\documentclass[reprint,
superscriptaddress,
showpacs,
nofootinbib,
prd,
aps,
floatfix,
]{revtex4-1}

\usepackage{amsmath,amssymb,amsfonts,amsthm}
\usepackage[english]{babel}
\usepackage{url}
\usepackage{bm}
\usepackage[latin1]{inputenc}
\usepackage{graphicx,rotating}
\usepackage[colorlinks=true,linkcolor=black, citecolor=blue]{hyperref}
\usepackage{csvsimple}
\usepackage{booktabs}
\usepackage{multirow, array}
\usepackage{slashed}

\usepackage{natbib}

\begin{document}
\title{Consistent description of angular correlations in $\beta$ decay for Beyond Standard Model physics searches}
\author{L. Hayen}
\email[Corresponding author: ]{lmhayen@ncsu.edu}
\author{A. R. Young}
\affiliation{Department of Physics, North Carolina State University, Raleigh, 27607 North Carolina, USA}
\affiliation{Triangle Universities Nuclear Laboratory, Durham, 27710 North Carolina, USA}

\date{\today}
\begin{abstract}
Measurements of angular correlations between initial and final particles in $\beta$ decay remain one of the most promising ways of probing the Standard Model and looking for new physics. As experiments reach unprecedented precision well into the per-mille regime, proper extraction of results requires one to take into account a great number of nuclear structure and radiative corrections in a procedure which becomes dependent upon the experimental geometry. We provide here a compilation and update of theoretical results which describe all corrections in the same conceptual framework, point out pitfalls and review the influence of the experimental geometry. Finally, we summarize the potential for new physics reach.
\end{abstract}

\maketitle


\section{Introduction}
Precision measurements of (nuclear) $\beta$ decay observables have to a significant extent defined the current status of the electroweak sector of the Standard Model \cite{Weinberg2009, Jensen2000, Sirlin2013, Donoghue1992}. Additionally, they are one of several promising and complementary pathways of finding and studying possible extensions to the latter in a theoretically relatively clean environment \cite{Holstein2014a}. In particular, due to the low energy transfers available in nuclear decays ($Q_\beta \lesssim 10$ MeV), many of the intricacies contained in the Standard Model are limited to higher-order effects and typically serve only to renormalize a number of coupling constants while leaving the bulk of the kinematic structure untouched. This is to the benefit of measurements of angular correlations between initial and/or final states as these are by definition relative effects, and generally do not require knowledge of all details of the decay distribution. Likewise, their measurement is experimentally promising as their relative nature allows for the cancellation of many otherwise dominant systematic uncertainties. Over the past decades, intense study in the neutron \cite{Mund2013, Markisch2019, Beck2019, Pattie2018} and mirror systems \cite{Severijns2008, Naviliat-Cuncic2009} have helped constrain and probe CKM universality and the presence of exotic scalar and tensor currents at a competitive level with those obtained from the LHC \cite{Naviliat-Cuncic2013, Wauters2014, Gonzalez-Alonso2018}. Already at the current experimental precision, however, several sources of theoretical higher-order input are required. We report here on a consistent description of the required corrections that experimental analyses need to take into account as the precision reaches and exceeds the per-mille level. While several of these results can be found in the literature, we argue that it is beneficial to put these results together in a comprehensive format as experimental analyses do not appear to be treated uniformly in the literature. This can lead to a incorrect comparison between different experimental results which in turn weakens their impact.

To leading order, exotic scalar or tensor currents in the weak interaction typically manifest themselves in the appearance of the so-called Fierz interference term, $b_F$. It modifies the total $\beta$ decay rate through a multiplicative factor
\begin{equation}
    \frac{d\Gamma}{dW_e} = \frac{d\Gamma_\text{SM}}{dW_e}\left[1 + b_F \frac{m_e}{W_e}\right],
    \label{eq:Fierz_appearance}
\end{equation}
where $d\Gamma_\text{SM}$ is the Standard model differential decay rate with $W_e$ the total $\beta$ particle energy. In the more modern language of $\beta$ decay effective field theories (EFT), it depends on new couplings $\epsilon$ according to \cite{Profumo2007, Gonzalez-Alonso2018, Cirigliano2013, Erler2005}
\begin{align}
    &b_F = \pm 2 \gamma \frac{1}{1+\rho^2} \nonumber \\
    &\times \text{Re}\left\{\frac{g_S\epsilon_S}{g_V(1+\epsilon_L+\epsilon_R)}+ \rho^2 \frac{4g_T\epsilon_T}{-g_A(1+\epsilon_L - \epsilon_R)} \right\},
    \label{eq:bF}
\end{align}
where the upper (lower) sign corresponds to $\beta^-$ ($\beta^+$) decay, $\gamma = \sqrt{1-(\alpha Z)^2}$, $g_A$ is defined as positive and all $\epsilon_i$ correspond to effective couplings arising due to new physics, with $\epsilon_i \sim (M_W/ \Lambda)^2$, with $M_W$ the mass of the $W$ boson and $\Lambda$ the scale of new physics. Per definition, $\Lambda \gg M_W$ and is typically at least of order TeV assuming naturalness arguments. The form factors are defined as $g_i = \langle p | \bar{u} O_i d | n \rangle$, where $g_S = 0.97(13)$ and $g_T = 0.987(55)$ are calculated on the lattice \cite{Gupta2018}.

Two correlations stand out from both an experimental and theoretical point of view. The first is the $\beta$-asymmetry ($A_\beta$), from which the discovery of parity violation was made \cite{Wu1957}, while the second is the $\beta$-$\nu$ ($a_{\beta \nu}$) correlation, which helped solidify the $V$-$A$ structure of the weak interaction \cite{Commins1983, Renton1990}. In an experimental setting one typically defines the differential decay rate according to their zeroth-order expressions
\begin{equation}
    \frac{d\Gamma_\text{SM}}{dW_ed\Omega_ed\Omega_\nu} = d\Gamma_0\left[1 + A_\beta P \beta \hat{J} \cdot \vec{p}_e  + a_{\beta \nu} \frac{\Vec{p}_e\cdot \Vec{p}_\nu}{W_eW_\nu}\right]
    \label{eq:exp_decay_rate}
\end{equation}
where the isotropic decay rate is  \cite{Hayen2018}
\begin{align}
    d \Gamma_0 &= \frac{G_F^2}{2\pi^3}F_0L_0C(W_e)g(W_e,W_0) \nonumber \\
    &\times K(W_e, W_0)p_eW_e(W_0-W_e)^2.
    \label{eq:gamma_0}
\end{align}
Here $\hat{J}$ is a unit vector along the initial polarization, $P = \langle M \rangle / J$ is the effective polarization, $W_e$ and $W_0$ are the $\beta$ particle total energy and endpoint energy in units of the electron rest mass, respectively, $\vec{p}_{e(\nu)}$ are the electron (antineutrino) three-momenta, and $\beta = p_e/W_e = v/c$ is the $\beta$-particle velocity. Additionally, $G_F \approx 10^{-5} m_p^{-2}$ is the Fermi constant, $F_0L_0$ is the Fermi function, $C(W)$ is the spin-independent shape factor, $g(W_e, W_0)$ is the well-known $\mathcal{O}(\alpha)$ energy-dependent radiative correction by Sirlin, $K(W_e, W_0)$ correspond to higher-order corrections of varying nature and $pW(W_0-W)^2$ is the phase space factors \cite{Hayen2018}. 

Naturally, correlations like those in Eq. (\ref{eq:exp_decay_rate}) contain different higher-order corrections due to their differing kinematic signature. As a result, Eq. (\ref{eq:exp_decay_rate}) describes effective correlations with several sub-dominant effects folded in. Some of these originate from nuclear structure, while others come about through electroweak radiative corrections or kinematic recoil. Additionally, like the appearance of the Fierz term, exotic scalar or tensor currents modify the effective values of $A_\beta$ and $a_{\beta \nu}$ \cite{Jackson1957a}. Both of these, however, depend only quadratically on exotic couplings so that their measurement obtains new physics sensitivity mainly from the appearance of the Fierz term. Because of this, measurements are typically interpreted in terms of an effective correlation (cf. Eq. (\ref{eq:Fierz_appearance}))
\begin{equation}
    \widetilde{X} = \frac{X}{1+b_F\langle m_e / W_e \rangle}
    \label{eq:coeff_X}
\end{equation}
where $X$ is any correlation coefficient, and $\langle m_e / W_e \rangle$ is the average of the $m_e / W_e$ term weighted by the spectrum (Eq. (\ref{eq:gamma_0})) over the experimental range. We already note that the validity of Eq. (\ref{eq:coeff_X}) depends on the experimental scheme, in particular for the $\beta$-$\nu$ correlation. This was the topic of Ref. \cite{Gonzalez-Alonso2016} and will be reiterated below.



The paper is organized as follows. In Sec. \ref{sec:general_expr} we provide the necessary theory input to calculate the Standard Model $\beta$ correlations to high precision, with corrections from nuclear structure, kinematics and radiative corrections. Sec. \ref{sec:mirror} discusses simplifications in mirror systems and the benefit of near-cancellation for sensitivity to $\rho$, the Fermi to Gamow-Teller mixing ratio. Precise measurements of the latter are additionally an ingredient in the determination of $|V_{ud}|$. Further, we discuss the way real experimental analyses are complicated due to a variety of effects in Sec. \ref{sec:exp_cond}. Finally, we discuss the new physics potential and sensitivity arising from precision measurements of $\beta$ decay correlations. We attach several appendices treating kinematic recoil corrections typically neglected in multipole formalisms and provide comparisons to other popular formalisms.

\section{General three-body decay rate}
\label{sec:general_expr}
We initiate our discussion through a definition of the general decay rate based on angular momentum conservation and the symmetries of the electroweak interaction. While several first-order expressions are available in the literature, in particular for the neutron \cite{Ando2004, Gardner2004, Gudkov2006}, we are mainly interested here in arbitrary spin changes. We first provide the general expression to lay the foundation for our discussion, and provide its rationale in the following section.

The general three-body $\beta$ decay rate summed over the helicities of the final states can then be written as
\begin{align}
    d\Gamma &= \frac{G_F^2}{2\pi^3}F_0L_0K(W_e, W_0) p_eW_e(W_0-W_e)^2 \nonumber \\
    \times &\left[f_0 +  \sum_{k \geq 1}f^{\beta\nu}_k P_k(\cos \theta_{\beta \nu}) + G_k(J_i)\biggl\{ f^{\sigma e}_k P_k(\cos \theta_e)\right. \nonumber \\
    &  + f^{\sigma \nu}_k P_k(\cos \theta_\nu) + f_k^{\sigma \times} P_k(\cos \theta_\times) \biggr\} \nonumber\\
    & + \text{higher orders }\Biggr]dW_e d\Omega_e d\Omega_\nu,
    \label{eq:general_decay_rate}
\end{align}
where $G_k$ is a polarization tensor of rank $k$ of the initial state, $P_k(\cos \theta)$ a Legendre polynomial of degree $k$, $f_0$ is the isotropic shape factor and $K(W_e, W_0)$ common corrections \cite{Hayen2018}. The `higher orders' in Eq. (\ref{eq:general_decay_rate}) stands for correlations involving more exotic combinations of momenta and higher powers (see, e.g., Ref. \cite{Ebel1957, Ivanov2013, Gudkov2006}), which we neglect here. The angles are defined as follows
\begin{align}
    \cos \theta_{\beta\nu} &= \frac{\vec{p}_e\cdot \vec{p}_\nu}{|\vec{p}_e||\vec{p}_\nu|}, \quad \cos \theta_e = \frac{\hat{J}\cdot \vec{p}_e}{|\vec{p}_e|}, \nonumber\\
    \cos \theta_\nu &= \frac{\hat{J}\cdot \vec{p}_\nu}{|\vec{p}_\nu|}, \quad \cos \theta_\times = \frac{\hat{J}\cdot (\vec{p}_e \times \vec{p}_\nu)}{|\vec{p}_e||\vec{p}_\nu|}.
\end{align}
Comparing to Eq. (\ref{eq:exp_decay_rate}), or more generally to the Jackson-Treiman-Wyld (JTW) categorization \cite{Jackson1957a}, one can recognize the usual asymmetries when limiting ourselves to $k=1$. Specifically, the ratio of spectral functions $f_k^i/f_0$ reduce to the well-known expressions of the $\beta$ correlations, with $f_1^{\beta \nu}/f_0$ the $\beta$-$\nu$ correlation ($a_{\beta \nu}$), $G_1 = \langle M \rangle / J = P$ and $f_1^{\sigma e}/f_0$, $f_1^{\sigma \nu}/f_0$, and $f_1^{\sigma \times}/f_0$ the $\beta$-asymmetry ($A_\beta$), $\nu$-asymmetry ($B_\nu$), and triple correlation ($D$), respectively. 

\subsection{Nuclear structure and kinematics}
All of the spectral functions of Eq. (\ref{eq:general_decay_rate}) depend on a combination of nuclear structure and QED corrections folded in together. Taking for now only the $\mathcal{O}(\alpha Z)$, low-energy part of the virtual photon exchange (i.e. the Coulomb interaction), the matrix element for $\beta$ decay can then be written down in a simple quantum mechanics picture with initial and final state interaction as \cite{Halpern1970, Behrens1982}
\begin{align}
    \mathcal{M} &= -2\pi i\delta(E_f-E_i) \langle f | T \left[\exp \left(-i\int_0^\infty dt\mathcal{H}^{Z_f}(t) \right) \right] \nonumber \\
    &\times \mathcal{H}_\beta(0)\, T \left[\exp \left(-i\int_{-\infty}^0 dt \mathcal{H}^{Z_i}(t) \right) \right] | i \rangle
    \label{eq:Halpern}
\end{align}
with $T$ the time-ordered product, $\mathcal{H}^Z$ the Hamiltonian density describing the Coulomb interaction and 
\begin{equation}
    \mathcal{H}_\beta(0) = \frac{G_F}{\sqrt{2}}V_{ud}H_\mu(0)L^\mu(0)
    \label{eq:H_current_current}
\end{equation}
is the Fermi current-current description of $\beta$ decay, with $L^\mu = \Bar{u}(p_e)\gamma^\mu(1-\gamma^5)v(p_\nu)$ the lepton current. Regardless of the description of the hadronic current, $H_\mu$, it is intuitively clear from Eq. (\ref{eq:Halpern}) that the final result depends on a convolution of the initial and final nuclear wave functions with the lepton current,all of which are modified because of the Coulomb interaction. Electroweak radiative corrections beyond the Coulomb interaction depend only at higher orders ($\mathcal{O}(\alpha^{n}Z^{n-1})$ for $n > 1$) on details of the nuclear wave functions \cite{Sirlin1986, Jaus1990, Jaus1987, Hayen2018, Hayen2019c}, and will be discussed in further detail in Sec. \ref{sec:radiative_corr} for angular correlations.

In the simplest case of the $J_i = 1/2$ to $J_f = 1/2$ transition of an elementary particle, such as the decay of the neutron, $H_\mu$ can be written down explicitly
\begin{align}
    H_\mu &= i \Bar{u}(p_f) \left\{ g_V\gamma_\mu + i\frac{\widetilde{g}_M}{2M}\sigma_{\mu \nu} q^\nu + \frac{\widetilde{g}_S}{2M}q_\mu  \right. \nonumber \\
    &\left. - g_A\gamma_\mu \gamma^5 + i\frac{\widetilde{g}_T}{2M}\sigma_{\mu \nu}q^\nu \gamma^5 + \frac{\widetilde{g}_P}{2M}q_\mu \gamma^5 \right\} u(p_i)
    \label{eq:neutron_nuclear_current}
\end{align}
from the requirement of Lorentz-invariance and initial and final spinors being on-shell. Here all $g_i(q^2)$ are dimensionless form factors and a function of $q^2 = (p_f - p_i)^2$, $\sigma_{\mu \nu} = \frac{i}{2}[\gamma_\mu, \gamma_\nu]$ and $M$ is the nucleon mass. The absence of second-class currents and the conserved vector current (CVC) hypothesis requires $\widetilde{g}_S = \widetilde{g}_T = 0$. Additionally, the application of CVC together with the Ademollo-Gatto theorem \cite{Ademollo1964} sets $g_V = 1$ up to corrections of $(q/M)^4$. CVC further allows for the interchange of weak and electromagnetic form factors. For example, for the neutron one can set $\widetilde{g}_M = \mu_p-\mu_n = 3.706$, where $\mu_{p,n}$ are the anomalous magnetic moment of proton and neutron, respectively. Finally, the partially conserved axial current relates $g_A$ and $\widetilde{g}_P$ through the Goldberger-Treiman relation
\begin{align}
    \widetilde{g}_P (q^2) = -g_A(0)\frac{(2M)^2}{m_\pi^2 - q^2}
    \label{eq:gP_goldberger_treiman}
\end{align}
assuming pion-pole dominance. In the limit of zero momentum transfer (as is appropriate in $\beta$ decay) one obtains $\widetilde{g}_P \approx -229$\footnote{While the large magnitude of $\widetilde{g}_P$ offsets somewhat the strong attenuation of the pseudoscalar matrix element, $\langle p | \gamma^5 | n \rangle \sim (v_n/c)^2$, its influence is felt only at the $10^{-4}$ level. In the case of strong cancellations, however, this can become relevant (see App. \ref{app:coefficients}).}. In this simple system then, one is left with only a single independent form factor, $g_A(q^2)$, to be determined either from experiment or from lattice QCD \cite{Chang2018, Gupta2018}.

In order to generalize Eq. (\ref{eq:neutron_nuclear_current}), several options have been explored in the literature. Almost all of these were developed half a century ago using the so-called \textit{elementary particle approach} \cite{Armstrong1972a, Holstein1974, Behrens1982}, using form factors coupled to specific angular momentum operators\footnote{Final expressions obtained this way are somewhat unsurprisingly very similar to those obtained by more modern EFT techniques. In that sense, using form factors near zero momentum with, e.g., a dipole formulation can be considered the phenomenological analog of separation of scales and the appearance of low-energy constants in current EFTs.}. Because of the small number of dominant operators in allowed $\beta$ decay, Holstein expanded the scalar product of Eq. (\ref{eq:H_current_current}) into a manifestly covariant form similar in spirit to Eq. (\ref{eq:neutron_nuclear_current}). While this has some clear advantages, it does not generalize well to arbitrary spin-parity changes and the Coulomb interaction has to be put in post-hoc. The second option takes inspiration from multipole decompositions in classical electrodynamics. Already in the 1960's it was shown that a Lorentz-invariant decomposition of a four-current exists \cite{Durand1962}, which becomes particularly simple in the Breit frame, where $\vec{p}_i = - \vec{p}_f$. The ramifications of choosing the Breit frame are discussed in Appendix \ref{app:kin_recoil}, and are related to the appearance of kinematical recoil corrections. The time component of Eq. (\ref{eq:neutron_nuclear_current}) can, e.g., be written as \cite{Stech1964}
\begin{align}
    H_0 = \sum_{LM}\mathcal{C}_{m_im_f;M}^{J_iJ_f;L}Y^M_L(\hat{q})\frac{(qR)^L}{(2l+1)!!}F_L(q^2),
    \label{eq:H_0_decomp}
\end{align}
where $\mathcal{C}$ contains a Wigner-$3j$ symbol, $Y_M^L$ is a spherical harmonic and $R$ is the nuclear radius so that $qR \ll 1$. Combining Eq. (\ref{eq:H_0_decomp}) with a multipole decomposition of the lepton current, $L_\mu$, the matrix element of Eq. (\ref{eq:Halpern}) can be calculated systematically for any spin-parity transition using spherical tensor algebra. As a consequence, in the analysis one never finds only a single term proportional to some $\cos \theta$, but rather a Legendre polynomial $P_l(\cos \theta)$, where each $l$ couples to a spherical tensor operator (and form factor) of rank $l$. The result is Eq. (\ref{eq:general_decay_rate}), where each spectral function $f_i$ is a combination of form factors.

In what follows, we will provide an outline of the calculation and report on the final results. All nuclear structure corrections were calculated in the Behrens-B\"uhring formalism \cite{Behrens1982} and reported here using a shorthand notation explained in Appendix \ref{app:notation_conventions}. Kinematic recoil corrections were obtained following the discussion in Appendix \ref{app:kin_recoil}. Sec. \ref{sec:mirror} discusses some qualitative results for experimentally interesting cases.

\subsubsection{Isotropic spectral function}

Since it normalizes the spectrum and thus appears in every $\beta$-correlation, we start with the isotropic spectral function, $f_0$. Together with the prefactors defined in Eq. (\ref{eq:general_decay_rate}), this is simply the $\beta$ spectrum when no other variables are measured. The isotropic spectral function was studied in great detail in Ref. \cite{Hayen2018} in the context of $\beta$-spectrum measurements aimed at directly measuring the Fierz term energy-dependence of Eq. (\ref{eq:Fierz_appearance}), and we simply write the general result
\begin{align}
    f_0 &= V_{0}^2\,{}^VC(W_e){}^VR_N(W_e, M) \nonumber \\
    &+ A_{10}^2\,{}^AC(W_e){}^AR_N(W_e, M),
    \label{eq:f_0}
\end{align}
where $C(W_e)$ is the so-called shape factor and $R_N$ captures the kinematic recoil corrections from a decay with nuclear mass $M$ (see also Appendix \ref{app:kin_recoil}). The dominant form factors in Eq. (\ref{eq:f_0}) reduce to the well-known expressions at zeroth order, with
\begin{subequations}
\begin{align}
    V_{0} &\equiv {}^VF_{000}(q^2 = 0) \simeq g_V \mathcal{M}_F \\
    A_{10} &\equiv {}^AF_{101}(q^2 = 0) \simeq -g_A \mathcal{M}_{GT}
\end{align}
\end{subequations}
where $\mathcal{M}_F$ ($\mathcal{M}_{GT}$) is the Fermi (Gamow-Teller) matrix element. Since the shape factor is $1+\mathcal{O}(10^{-2})$, and the kinematic recoil corrections are at most $\mathcal{O}(10^{-3})$, to leading order (LO) we have
\begin{equation}
    f_0^{LO} = |g_V \mathcal{M_F}|^2 + |g_A\mathcal{M}_{GT}|^2 + \mathcal{O}(10^{-2}),
    \label{eq:f_0_LO}
\end{equation}
with the percent-order corrections arising from finite size corrections and induced currents \cite{Hayen2018}.

\subsubsection{$\beta$-Asymmetry}
Following the discussion on Eqs. (\ref{eq:exp_decay_rate}) and (\ref{eq:general_decay_rate}), we define an effective $\beta$-asymmetry according to
\begin{equation}
    A_\beta P \beta \cos \theta_e \cong \sum_{k \geq 1} G_k(J_i)\frac{f_k^{\sigma e}}{f_0} P_k(\cos \theta_e),
    \label{eq:A_equiv}
\end{equation}
where we note an approximate equivalence to stress the fact that the angular structure is different in both sides. In the simplest case where $k=1$, the polarization tensor is simply $G_1 = \langle M_i \rangle / J_i$ and $P_1(\cos \theta_e) = \cos \theta_e$. After some tedious algebra we find
\begin{align}
    f^{\sigma e}_1 &= \sqrt{\frac{6J_i}{J_i+1}} \Lambda_1(W_e) \beta  \left[\mathcal{S}_1A_{10}^2 \pm \sqrt{2/3}\,V_{0}A_{10}\right. \nonumber \\
     &\left. + \alpha_1^0 + \alpha_1^1 W_e + \alpha_1^2 W_e^2 \right]
     \label{eq:f_sigma_e}
\end{align}
where the upper (lower) sign refers to $\beta^-$ ($\beta^+$), $\mathcal{S}_1$ is a spin-coupling coefficient, all the $\alpha_1^i$ are of order $\mathcal{O}(10^{-2})$ and are listed in full in Appendix \ref{app:coefficients}. The factor $\Lambda_1(W_e)$ in Eq. (\ref{eq:f_sigma_e}) is a type of Coulomb function originally defined by Behrens and J\"anecke \cite{Behrens1969} which is $\mathcal{O}(1)$ and not present in most other formalisms \cite{Holstein1974}. It was calculated numerically long ago \cite{Behrens1969}, and shows deviations from unity only at the few $10^{-4}$ level for energies below a few MeV. At the current and future level of precision, however, its influence can already be felt.

Analogous to Eq. (\ref{eq:f_0_LO}), we can write $f_1^{\sigma e}$ to leading order to find
\begin{equation}
    f_1^{\sigma e} \sim \mp\sqrt{\frac{6J_i}{J_i+1}}\beta \left(\mathcal{S}_1 A_{10}^2 \pm \sqrt{2/3}V_0A_{10} \right) + \mathcal{O}(10^{-2}).
    \label{eq:f_1_sigma_e_LO}
\end{equation}
Extracting now a factor $V_0^2$ from both $f_0$ and $f_1^{\sigma e}$ and defining the Fermi to Gamow-Teller mixing ratio (see Sec. \ref{sec:mirror} and Appendix \ref{app:coefficients})
\begin{align}
    \rho &\equiv \frac{A_{10}}{V_0} \nonumber \\
    &\simeq -\frac{g_A \mathcal{M}_{GT}}{g_V\mathcal{M}_F},
\end{align}
(taking $g_A$ positive as before) we recover the usual leading-order result for the $\beta$-asymmetry \cite{Severijns2006},
\begin{equation}
    A_{\beta}^{LO} = \mp \sqrt{\frac{6J_i}{J_i+1}}\frac{\mathcal{S}_1\rho^2 \pm \sqrt{2/3}\,\rho}{1+\rho^2}.
    \label{eq:A_LO}
\end{equation}
For $J\to J$ transitions we have $\mathcal{S}_{1} = \{6J(J+1)\}^{-1/2}$ (Appendix \ref{app:coefficients}), so that for $J=1/2$ transitions one recovers the usual result,
\begin{equation}
    A_\beta^{LO} \stackrel{J=1/2}{=} \mp \frac{2}{\sqrt{3}}\frac{\rho^2/\sqrt{3}\pm \rho}{1+\rho^2}.
\end{equation}

The next spectral function, $f_2^{\sigma e}$, typically denoted by the `anisotropy', can similarly be calculated. Following the spherical tensor algebra, it couples with the polarization tensor of rank 2
\begin{equation}
    G_2(J_i) = \frac{1}{J_i^2}\left[\langle M_i^2 \rangle - \frac{1}{3}J_i(J_i+1) \right].
\end{equation}
The anisotropy is then
\begin{equation}
    f_2^{\sigma e} = (p_e R) \beta \nu_{12} \mathcal{S}_2 \left[\alpha_2^0 + \alpha_2^1 W_e \right]
    \label{eq:f_2_sigma_e}
\end{equation}
where once again all $\alpha_2^i$ are $\mathcal{O}(10^{-2})$, and $\nu_{12} = 1 + \mathcal{O}\{(\alpha Z)^2\}$ another Coulomb function \cite{Behrens1969, Behrens1982}. The prefactor $(p_e R) \lesssim 0.1$ sets the overall magnitude of the correction as being at most a few $10^{-3}$ effect depending on $G_2$. Since the Legendre polynomial is, however,
\begin{align}
    P_2(\cos \theta_e) = \frac{1}{2}(3 \cos^2 \theta_e - 1)
    \label{eq:legendre_P2}
\end{align}
the influence of $f_2^{\sigma e}$ depends on potential cancellations in $f_1^{\sigma e}$ and on the experimental geometry and solid angle. We will get back to this in Sec. \ref{sec:mirror} and \ref{sec:exp_cond}.

\subsubsection{$\beta$-$\nu$ Correlation}
Analogously to Eq. (\ref{eq:A_equiv}) we define the effective $\beta$-$\nu$ correlation coefficient as
\begin{equation}
    a_{\beta \nu} \beta \cos \theta_{\beta \nu} \cong \sum_{k \geq 1} \frac{f_k^{\beta \nu}}{f_0}P_k(\cos \theta_{\beta \nu})
\end{equation}
where like the $\beta$-asymmetry the r.h.s. has a richer structure than the traditional l.h.s. The first order result for $k=1$ can similarly be found
\begin{equation}
    f_1^{\beta\nu} = \beta\Lambda_1 \left[V_{0}^2 - \frac{1}{3}A_{10}^2 + \widetilde{\alpha}_1^0 + \widetilde{\alpha}_1^1 W + \widetilde{\alpha}_1^2 W^2 \right]
    \label{eq:f_1_beta_nu}
\end{equation}
where it is well-known that no vector-axial vector cross terms appear like in Eq. (\ref{eq:f_sigma_e}). Likewise, all $\widetilde{\alpha}_1^i$ are $\mathcal{O}(10^{-2})$ and are listed in Appendix \ref{app:coefficients}. In the same spirit, it is well-known \cite{Holstein1974} that $\widetilde{\alpha}_1^i$ contains a smaller set of induced currents than, for example, $\alpha_1^0$, such as the so-called induced tensor form factor.

Similarly to Eqs. (\ref{eq:f_0_LO}) and (\ref{eq:f_1_sigma_e_LO}), the LO behavior of $f_1^{\beta\nu}$ is
\begin{equation}
    f_1^{\beta\nu} \sim \beta \left(|g_V\mathcal{M}_F|^2 - \frac{1}{3}|g_A\mathcal{M}_{GT}|^2\right) + \mathcal{O}(10^{-2}),
\end{equation}
so that the LO $\beta$-$\nu$ asymmetry is the well-known expression
\begin{equation}
    a_{\beta\nu}^{LO} = \frac{1-\frac{1}{3}\rho^2}{1+\rho^2}.
\end{equation}

The anisotropy in the $\beta$-$\nu$ correlation can likewise be calculated
\begin{equation}
    f_2^{\beta\nu} = \beta (p_e R) [\widetilde{\alpha}_2^0 + \widetilde{\alpha}_2^1 W_e]
    \label{eq:f_2_beta_nu}
\end{equation}
and where once again all $\widetilde{\alpha}_2^i$ are of order $\mathcal{O}(10^{-2})$, making this at most a few $10^{-3}$ effect with the same angular structure as Eq. (\ref{eq:legendre_P2}) and sensitivity as discussed before for the $\beta$-asymmetry. Note that from symmetry requirements, $f_2^{\beta\nu}$ contains no nuclear form factors, and is instead only a kinematic feature arising from the three-body decay (see also Ref. \cite{Gudkov2006}).

\subsection{Radiative corrections}
\label{sec:radiative_corr}
In using Eq. (\ref{eq:Halpern}) we have only taken into account the Coulomb interaction, i.e., the large-wavelength behaviour of the of virtual photon exchanges between initial and final states. That is not the only $\mathcal{O}(\alpha)$ correction that shows up, however, which are more generally known as electroweak radiative corrections. The topic of radiative corrections has a rich history which lies at the heart of our current understanding of electroweak interactions and the Standard Model, and has been reviewed in several excellent works \cite{Sirlin2013, Towner1994}. Instead, we shall again be brief, and only summarize results available in the literature.

The order $\alpha$ photonic radiative corrections are the result of three processes: ($i$) virtual photon exchange between initial and final states, ($ii$) real photon emission from external lines, and ($iii$) wave function renormalization of the external legs. The results of these processes have typically been calculated in the way proposed by Sirlin \cite{Sirlin1967}, with a separation of the processes according to photon momentum. It was shown that a relatively clean, gauge-invariant separation could be obtained between contributions for high photon momentum ($k \gg p_e$), resulting in a renormalization of the coupling constants \cite{Sirlin1967}
\begin{subequations}
\begin{align}
    g_V &\to g_V^\prime \equiv g_V \left(1 + \frac{\alpha}{2\pi}c \right) \label{eq:inner_RC_gV}\\  
    g_A &\to g_A^\prime \equiv g_A \left(1 + \frac{\alpha}{2\pi}d \right)
    \label{eq:inner_RC_gA}
\end{align}
\end{subequations}
(known as the \textit{inner} radiative correction, $\Delta_R^{V(A)} = \alpha/\pi c(d)$ \cite{Seng2018, Hayen2019c, HayenGTRC}) and those at low photon momentum ($k \leq p_e$) with a dependence on final state kinematics (known as the \textit{outer} radiative corrections, $\delta_R(W_e)$) \cite{Sirlin1967, Jaus1970, Hayen2018}. This is possible because results are dominated by either infrared divergences ($\delta_R$) or high-energy ($\gg m_e$) electroweak and strong physics\footnote{The notable exception is, of course, the $\gamma W$ box which is also sensitive to physics at the nuclear scale. For the purpose of this discussion, however, we consider it fully part of $\Delta_R$ (see Ref. \cite{Gorchtein2018}).} ($\Delta_R
^{V,A}$). Practically all other calculations have been constructed in the same way \cite{Shann1971, Garcia1982, Toth1986, Gluck1997}.

In terms of these renormalized coupling constants, the kinematic structure of the radiative corrections can be written down for the lowest order results
\begin{align}
    d\Gamma &\approx d\Gamma_0 \Biggl[f_0^\prime \left(1 + \frac{\alpha}{2\pi}g \right) +  f^{\beta\nu\prime}_1 \left(1 + \frac{\alpha}{2\pi}h \right)\cos \theta_{\beta \nu}\nonumber \\
    &+ P\biggl\{ f^{\sigma e\prime}_1 \left(1 + \frac{\alpha}{2\pi}h \right) \cos \theta_e + f^{\sigma \nu\prime}_1 \left(1 + \frac{\alpha}{2\pi}g \right) \cos \theta_\nu \nonumber \\
    &+ f_1^{\sigma \times\prime} \cos \theta_\times\biggr\}+ \sum_{k \geq 2} f_k^{i\prime}\Biggr]dW_e d\Omega_e d\Omega_\nu
    \label{eq:decay_rate_RC}
\end{align}
where all primed $f_k^\prime$ correspond to the usual expressions, but using the renormalized coupling constants of Eqs. (\ref{eq:inner_RC_gV}) and (\ref{eq:inner_RC_gA}), and the outer radiative corrections are well known \cite{Shann1971, Garcia1982}
\begin{align}
    &g(W_e, W_0) = 3 \log M_p - \frac{3}{4} + \frac{4}{\beta}L\left(\frac{2\beta}{1+\beta} \right) \nonumber \\
    &+ 4 \left(\frac{\tanh^{-1}\beta}{\beta}-1\right)\left[\frac{W_0-W_e}{3W_e} - \frac{3}{2} + \ln[2(W_0-W_e)] \right] \nonumber \\
    & + \frac{\tanh^{-1}\beta}{\beta} \left[2(1+\beta^2) + \frac{(W_0-W_e)^2}{6W_e^2} - 4 \tanh^{-1}\beta \right]
    \label{eq:g_sirlin}
\end{align}
and 
\begin{align}
    &h(W_e, W_0) = 3 \log M_p - \frac{3}{4} + \frac{4}{\beta} L\left(\frac{2\beta}{1+\beta} \right) \nonumber \\
    & + 4 \left(\frac{\tanh^{-1} \beta}{\beta} - 1\right) \left[ \ln[ 2(W_0-W_e)] - \frac{3}{2} + \frac{W_0-W_e}{3W_e\beta^2} \right. \nonumber \\
    &\left. + \frac{(W_0-W_e)^2}{24 W_e^2 \beta^2}\right] + \frac{4}{\beta} \tanh^{-1} \beta (1-\tanh^{-1} \beta)
    \label{eq:h_shann}
\end{align}
where $L(x) = \int_0^x (\log (1-t)/t) dt$ is the Spence function and $M_p$ is the mass of the proton. It is well-known that the triple correlation, $f_1^{\sigma \times}$, contains no outer radiative corrections \cite{Gluck1990}. We have neglected additional outer radiative corrections to higher-order terms, as they would constitute only a $\mathcal{O}(\alpha/2\pi) \sim 10^{-3}$ shift on top of an already small effect (see Eqs. (\ref{eq:f_2_sigma_e}) and (\ref{eq:f_2_beta_nu})), but for consistency treat them using the renormalized coupling constants.

The measurement of a $\beta$ correlation has the particular advantage of being sensitive only to the relative differences between the isotropic and correlation spectral functions. As a consequence, at first sight there is no additional kinematic structure arising from radiative corrections to the $\nu$-asymmetry ($B_\nu$), while the $\beta$-asymmetry ($A_\beta$) and $\beta$-$\nu$ correlation ($a_{\beta\nu}$) are modified by
\begin{align}
    R &\equiv \frac{1+\frac{\alpha}{2\pi}h}{1+\frac{\alpha}{2\pi}g}\nonumber \\
    &\approx 1+ \frac{\alpha}{2\pi} \left\{(1-\beta^2)\frac{4(W_0-W_e)}{3W_e\beta^2}\left(\beta^{-1}\tanh^{-1} \beta -1 \right)\right.\nonumber \\
    &+ \frac{(W_0-W_e)^2}{6W_e^2\beta^2}\left[(1-\beta^2)\beta^{-1}\tanh^{-1} - 1 \right] \nonumber \\
    &+ 2(1-\beta^2)\beta^{-1}\tanh^{-1}\beta \Biggr\}.
    \label{eq:R_approx}
\end{align}
It is interesting to note that as $\beta \to 1$ the difference between $g$ and $h$ reduces to a single term in the second line. In the case of the neutron, this has been numerically estimated by Fukugita and Kubota \cite{Fukugita2004}
\begin{equation}
    R \approx 1 +\left(- 1.63 + 4.11W_e^{-1} + 0.236\, W_e\right) \cdot 10^{-3}.
\end{equation}

\begin{figure}[h!]
    \centering
    \includegraphics[width=0.48\textwidth]{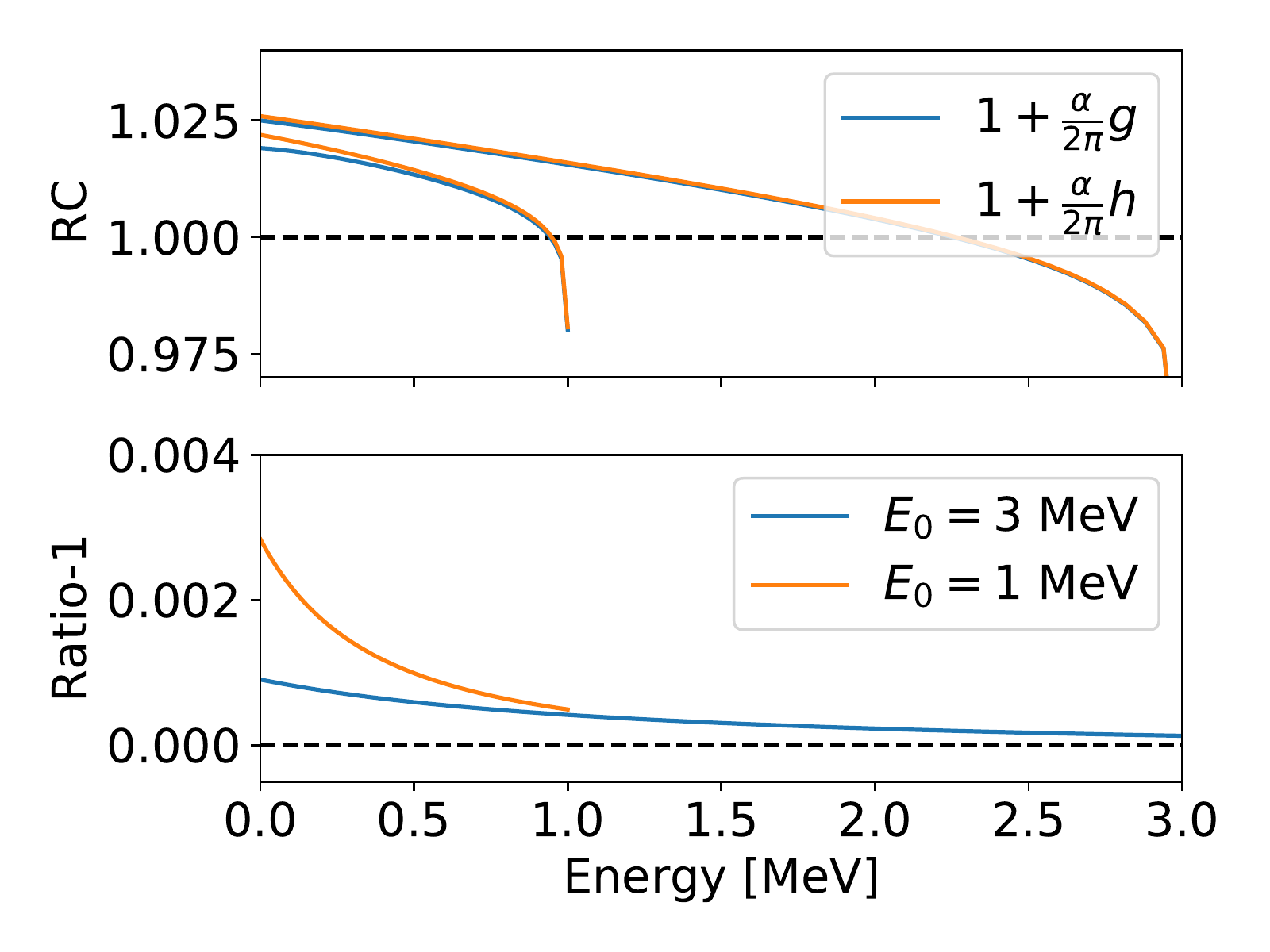}
    \caption{Radiative corrections due to $g$ (Eq. (\ref{eq:g_sirlin})), $h$ (Eq. (\ref{eq:h_shann})) and their ratio, $R$ (Eq. (\ref{eq:R_approx})) for two different endpoint energies.}
    \label{fig:RC}
\end{figure}

Figure \ref{fig:RC} shows the different radiative corrections and their ratio for two $\beta$ transitions with a 1 and 3 MeV endpoint. Interesting to note that is that $R > 1$ and the slope becomes stronger at low energy for lower endpoint energies. Finally, the limiting $\beta \to 1$ behavior coincides with the expectations.

Experimentally, however, not all of these results are applicable. Whenever a correlation with the (anti)neutrino is measured, the situation is not as clear-cut as one typically measures the nuclear recoil rather than the emitted (anti)neutrino. As a consequence, for an extraction of $a_{\beta\nu}$ and $B_\nu$ from experiment it is typically not appropriate to use the above expressions. We will discuss this further in Sec. \ref{sec:exp_cond}.

\section{Mirror decays as a testing ground}
\label{sec:mirror}
The previous section summarized the required theoretical input arising from nuclear structure and radiative corrections. In all measurements of aforementioned correlations the largest unconstrained parameter for a mixed ($J \to J$, $J > 0$) is the mixing ratio, $\rho$, traditionally defined as \cite{Severijns2008, Naviliat-Cuncic2009}
\begin{align}
    \rho &= \frac{A_{10}}{V_{0}}\left[\frac{1+\Delta_R^A}{1+\Delta_R^V} \right]^{1/2} \nonumber \\
    &\approx \frac{g_A\mathcal{M}_{GT}^0}{g_V\mathcal{M}_F^0}\left[\frac{(1+\delta_{NS}^A-\delta_C^A)(1+\Delta^A_R)}{(1+\delta_{NS}^V-\delta_C^V)(1+\Delta^V_R)} \right]^{1/2}
    \label{eq:rho_long}
\end{align}
where $\delta_{NS(C)}$ are nuclear structure and isospin breaking corrections to the Fermi ($F$) and Gamow-Teller ($GT$) matrix elements in the limit of isospin symmetry, denoted by the $``0"$ superscript. \cite{Towner2010}. The latter are typically assumed to be equal for axial and vector parts, meaning experiments measure the ratio of many-body matrix elements and renormalized coupling constants. Note that we have also neglected the presence of second-class currents here, which are briefly discussed in Appendix \ref{app:coefficients}.

An accurate determination of $\rho$ is extremely interesting from a physics point of view (see Sec. \ref{sec:new_physics_sensitivity}), however the precision that can be obtained depends both on the sensitivity of the correlation coefficient to $\rho$ and the uncertainty on the remaining theory input. In both cases, so-called isospin $T=1/2$ mirror $\beta$ decays are a prime candidate \cite{Severijns2008, Naviliat-Cuncic2009}.

It is important to note, however, that the notation in Eq. (\ref{eq:rho_long}) can be somewhat deceiving. The reason for the separation of $\mathcal{M}_F$ into $\mathcal{M}_F^0$ and isospin breaking corrections is because CVC and isospin symmetry allow for a certain determination of $\mathcal{M}_F^0$, with the former additionally guaranteeing that no additional corrections appear beyond the impulse approximation result. The leading Gamow-Teller form factor, however, is very different. Not only is $\mathcal{M}_{GT}^0$ not determined by any symmetry, the absence of the conservation of the axial current means additional contributions beyond the impulse approximation result necessarily enter, traditionally denoted by core-polarization and meson-eschange effects. Phenomenologically, this is often obfuscated by a so-called quenching factor to the axial vector coupling constant \cite{Wilkinson1973, Towner1994, Suhonen2017a}. As ab initio calculations ramp up their capabilities in this regard \cite{Gysbers2019}, however, using more sophisticated ways of solving the $N$-body Schr\"odinger equation, a more correct way of presenting $\rho$ would be
\begin{equation}
    \rho \approx \frac{g_A^{QCD}\mathcal{F}_{GT}(0)}{g_V\mathcal{M}_F^0}\left[\frac{1 + \Delta_R^A-\Delta_R^ V}{1+\delta_{NS}^V-\delta_C^V} \right]^{1/2}
\end{equation}
where $g_A^{QCD}$ is the renormalized value solely due to strong interaction effects, and $\mathcal{F}_{GT}(0)$ is the normalized nuclear response to a nucleonic Gamow-Teller operator near zero momentum transfer. If $g_A$ is instead taken from an experimental measurement in the neutron
\begin{equation}
    g_A^n = g_A^{QCD}\left[1+\,^{n}\Delta_R^A-\,{}^n\Delta_R^V\right]^{1/2},
\end{equation}
a partial cancellation occurs with the inner radiative corrections to the mirror Gamow-Teller transition, which are recently found to contain transition-dependent terms \cite{Hayen2019c, Gorchtein2018}. Note that all of these effects require and in part originate from an internally consistent set of definitions used both in experimental extraction and theoretical analysis \cite{Hayen2019c}. With the isospin-breaking corrections being an $0.2\%$ to $1\%$ effect \cite{Naviliat-Cuncic2009}, and differences between $\Delta_R^A$ and $\Delta_R^V$ on the $10^{-3}$ level \cite{Hayen2019c, Gorchtein2018}, such differences become relevant in the neutron and low-mass systems.

\subsection{Cancellation for precision}
Looking at Eqs. (\ref{eq:f_sigma_e}) and (\ref{eq:f_1_beta_nu}) there is a potential for cancellation between the two main terms for a mixed decay. In particular, when
\begin{equation}
    \rho^2 \approx \left\{
    \begin{array}{lc}
    \mp \rho\sqrt{2/3}\, \mathcal{S}_1^{-1} & (A_\beta) \\
        3 & (a_{\beta \nu})
    \end{array}
    \right.
    \label{eq:rho_cancellation}
\end{equation}
significant cancellation occurs. This is interesting since typically the values of correlation coefficients are very sensitive to the value of $\rho$ near such a turnover point. 

\begin{figure}[h!]
    \centering
    \includegraphics[width = 0.48\textwidth]{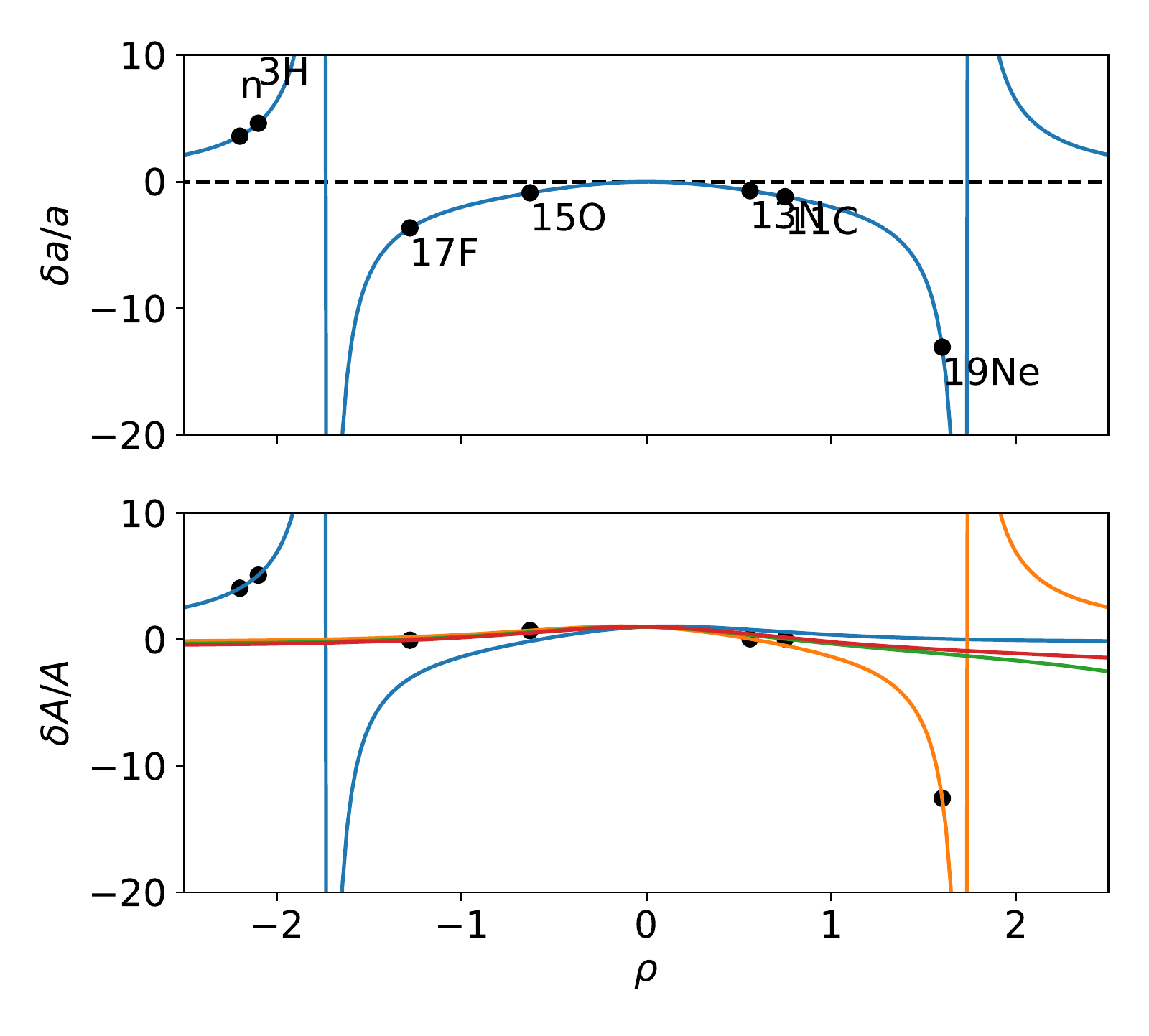}
    \caption{(Top) Calculated sensitivities to $\delta \rho / \rho$ from $\delta a / a$. The sensitivity is symmetric w.r.t. $\rho$ and spin-independent. (Bottom) Calculated sensitivities to $\delta \rho / \rho$ from $\delta A / A$ for $\beta^-, J = 1/2$ (blue), $\beta^+, J=1/2$ (orange), $\beta^+, J=3/2$ (green), $\beta^+, J = 5/2$ (red).}
    \label{fig:sensitivities}
\end{figure}
For $J \to J$ transitions like nuclear mirrors, $(\mathcal{S}_1)^{-1} = \sqrt{6J(J+1)}$ (see Appendix \ref{app:coefficients}). The neutron, for example, has $\rho = g_A \sqrt{3}$ and both $A_\beta$ and $a_{\beta \nu}$ are close to cancellation, since $(\mathcal{S}_1)^{-1} = 3/\sqrt{2}$. In this case, one finds $\delta A/A \approx 4.0 \delta \rho / \rho$ and $\delta a/a \approx 3.6 \delta \rho / \rho$. This results in an enhancement factor on $\rho$ of a factor 4, which is of particular interest for $V_{ud}$ and CKM unitarity tests discussed in Sec. \ref{sec:new_physics_sensitivity}. Table \ref{tab:sensitivity_mirrors} shows the enhancement factor for all nuclear mirrors up to mass 19, where advances in nuclear \textit{ab initio} theory are also likely to make significant progress in the near future \cite{Cirgiliano2019}.

\begin{table}[h!]
    \centering
    \begin{ruledtabular}
    \begin{tabular}{c|rrrrrrr}
    Nucleus & n & $^3$H & $^{11}$C & $^{13}$N & $^{15}$O & $^{17}$F & $^{19}$Ne \\
    \hline
    $\rho$ & $-2.20$ & $-2.10$ & $0.75$ & $0.56$ & $-0.63$ & $-1.28$ & $1.60$ \\
    $J$ & $1/2$ & $1/2$ & $3/2$ & $1/2$ & $1/2$ & $5/2$ & $1/2$ \\
    $\delta A_\beta / A_\beta$ & 4.0 & 5.1 & 0.04 & 0.04 & 0.7 & $-0.06$ & $-12.6$\\
    $\delta a_{\beta\nu}/a_{\beta\nu}$ & 3.6 & 4.6 & $-1.2$ & $-0.7$ & $-0.9$ & $-3.6$ & $-13.1$\\
    \end{tabular}
    \end{ruledtabular}
    \caption{Calculated sensitivities to $\delta \rho / \rho$ for the lowest mass mirrors, with approximate $\rho$ values taken from \cite{Severijns2008} and the leading order expressions.}
    \label{tab:sensitivity_mirrors}
\end{table}

As expected from Eq. (\ref{eq:rho_cancellation}), mirrors with $|\rho| \sim \sqrt{3}$ show strong enhancement factors, making them experimentally interesting candidates. From Fig. \ref{fig:sensitivities} it is also clear that for mirrors with spins higher than $J = 1/2$, the largest sensitivity to $\rho$ is likely to come from a measurement of the $\beta$-$\nu$ correlation. A turning point sensitivity is reached for $\rho = 2/3$, where $\delta a / a = \delta \rho / \rho$.

A clear disadvantage of such a cancellation, however, is that as leading order effects become small, initially subdominant corrections gain in relative importance. Taking $^{19}$Ne as an example, since its leading order $\beta$-asymmetry is about -4\% \cite{Calaprice1975a}, the relative importance of all subdominant corrections is now enlarged by a factor 25, which puts more stringent constraints on additional theory input.

\subsection{Remaining theory uncertainty}
Regardless of a potential cancellation in any of the coefficients, the experimental precision is such that in any case subdominant effects must be taken into account to varying degree. These have been summarized in the previous section, with additional complications due to the experimental geometry and detection scheme treated in the following section.

From a theory point of view, the precision bottleneck lies in the accurate calculation of nuclear matrix elements, in particular those stemming from induced currents. Because the $\beta$ decay occurs within an isospin multiplet, however, mirrors have a distinct advantage. Due to the conserved vector current, all vector form factors can be determined exactly in the limit of isospin symmetry. The Fermi matrix element, $\mathcal{M}_F$, is equal to unity for $T=1/2$, with isospin breaking corrections calculated in a many-body code (see Eq. (\ref{eq:rho_long})) \cite{Towner2010}. As mentioned before, all induced scalar form factors are identically equal to zero. Further, the invocation of CVC trivializes most of the additional theory input, as most recoil form factors are either zero or known to very high precision. This is the case for the so-called 'weak magnetism' form factor, $V_{11}$ ($\sim b(q^2)$ in Holstein's notation, see Appendix \ref{app:comparison}), which can be related to the isovector magnetic moment of initial and final states for $T=1/2$ or to the $M1$ decay width of the corresponding $\gamma$ transition for $T=1$. Finally, the first-class part of the induced tensor form factor, $A_{1}$ ($\sim d(q^2)$ in Holstein's notation, see Appendix \ref{app:comparison}), is identically equal to zero within an isospin multiplet. 

Besides the mixing ratio, $\rho$, this leaves at least two more subdominant sources of nuclear structure input, since axial form factors are not protected by any symmetry. In particular, the so-called induced pseudoscalar coupling, $A_{12}$ ($\sim h(q^2)$ in Holstein's notation, see Appendix \ref{app:comparison}), must be calculated by a many-body method unless it is trivially equal to zero \cite{Behrens1978, Hayen2018}. Higher-order form factors such as $A_{22}$ can additionally contribute for transitions with $J \geq 1$, and must be calculated using many-body methods. From the expressions in the appendix, one can estimate their influence to be at the few $10^{-4}$ level. Finally, there is an induced pseudoscalar contribution proportional to $\widetilde{g}_P$ (see Eq. (\ref{eq:gP_goldberger_treiman})), which is discussed to some depth in Ref. \cite{Hayen2018} and the appendix, and can also contribute up to the $10^{-4}$ level.


\section{Experimental conditions}
\label{sec:exp_cond}
Section \ref{sec:general_expr} contained some foreshadowing and caveats concerning the validity of the equations presented or the conclusions taken from it. The formulae written above correspond to an ideal situation, i.e., a perfect cancellation of all terms but the one of interest, $4\pi$ solid angle, measurement of the (anti)neutrino rather than the recoiling nucleus, perfect energy measurements, and so on. 

An analysis attempting an extraction of the correct quantity runs into at least three conceptual difficulties due to experimental conditions: ($i$) relative rate measurements in an open geometry folds in other observables and higher-order polarization effects ($ii$) real photons in radiative $\beta$ decay change the kinematics and must be accounted for ($iii$) a measurement of a correlation may not allow for the effective parametrization $\widetilde{X}$ (Eq. (\ref{eq:coeff_X})) \cite{Gonzalez-Alonso2016}. Besides this, several systematic effects emerge related to detector performance, e.g., through linearity and efficiency. Finally, measurements not relying on initial polarization can contain contamination from experimental residual polarization which may not be known to great precision.

Additional complication arises because of the experimental scheme and which final states are detected. Even though any three body decay allows for only two independent degrees of freedom, several combinations are typically used in the literature. Due to the additional richness it brings, several modern experiments measure, e.g., both the $\beta$ particle and the recoiling nucleus. This opens up the Dalitz distribution for analysis
\begin{align}
    \frac{d\Gamma}{dW_edW_f} &= \frac{G_F^2}{4\pi^3}W_fW_eq \nonumber \\
    &\times \left\{1+b_F \frac{1}{W_e} + \frac{1}{F_0}\sum_k F^{\beta\nu}_kP_k(\cos \theta_{\beta\nu})\right\},
    \label{eq:dalitz}
\end{align}
where all $F_k$ correspond to the modified spectral functions due to radiative corrections, e.g., $F_0 = f_0^\prime [1+(\alpha/2\pi)g]$, for brevity. The $\beta$-$\nu$ angle is then simply
\begin{equation}
    \cos \theta_{\beta\nu}  = \frac{p_f^2-p_e^2-q^2}{2W_eq}
    \label{eq:cos_beta_nu_recoil}
\end{equation}
where $q = W_0-W_e-W_f$ is the antineutrino energy and $W_f$ is the recoil energy (neglecting the real photon momentum).

The following sections summarize results arising from higher-order effects and discuss the complications due to real photons and the detection scheme on Standard Model comparisons. Following the discussion in Sec. \ref{sec:mirror}, these argument become particularly relevant in the case of strong cancellations such as several mirror systems.

\subsection{Solid angle}
\label{sec:solid_angle}
In a typical experiment one measures the difference in integrated count rates
\begin{equation}
    X = \frac{N^\uparrow - N^\downarrow}{N^\uparrow + N^\downarrow},
    \label{eq:exp_def_X}
\end{equation}
or with some more complicated super-ratio, where $N^{\uparrow (\downarrow)}$ are integrated count rates either in separate (usually opposite w.r.t. the maximum of $X$) or a single detector and instead changing, e.g., the polarization direction. As a consequence, everything in Eq. (\ref{eq:general_decay_rate}) besides $X$ also folds into $N$, where now the residual effect depends on the experimental conditions and geometry. 

The full decay rate results from an integration over all remaining variables of Eq. (\ref{eq:general_decay_rate})
\begin{align}
    \Gamma = &\frac{1}{(4\pi)^2}\int_1^{W_0}dW_e d\Gamma_0 F_0 \int_{-1}^1 d\cos \theta_e \int_{0}^{2\pi}d\phi_e \nonumber \\
    &\times\int_{-1}^1d\cos \theta_\nu \int_{0}^{2\pi} d\phi_\nu\, \mathcal{D}
    \label{eq:decay_rate_integration}
\end{align}
where the $z$-axis is along the initial polarization if present and random otherwise and
\begin{align}
    \mathcal{D} &= 1 + b_F \frac{1}{W_e} + \frac{1}{F_0} \sum_{k \geq 1} F_k^{\beta\nu}P_k(\cos \theta_{\beta\nu})\nonumber \\
    &+ G_k(J_i) \biggl\{F_k^{\sigma e}P_k(\cos \theta_e) + F_k^{\sigma \nu}P_k(\cos \theta_\nu) \nonumber \\
    &+ F_k^{\sigma\times}P_k(\cos \theta_\times) \biggr\}.
\end{align}
In practice, the angular integration limits depend on the experimental geometry and the energy integration requires a convolution with a calibrated detector response function, e.g., $\mathcal{R}(W_e, E)$ as the probability of measuring $E$ for a $\beta$ particle with real energy $W_e$, to find\footnote{In general the detector response function depends not only on the particle energy but also, e.g., on its angle of incidence into the detector face.}
\begin{align}
    \Gamma_\text{exp} &= \frac{1}{(4\pi)^2} \int_{E^\text{min}}^{E^\text{max}}dE\int_{1}^{W_0}dW_e\mathcal{R}(W_e, E)d\Gamma_0 F_0 \nonumber \\
    &\times \int_{\Omega_e^\text{exp}}d\Omega_e\mathcal{E}(\Omega_e)\int_{\Omega_\nu^\text{exp}}d\Omega_\nu\mathcal{E}(\Omega_\nu)\, \mathcal{D}.
    \label{eq:exp_rate_integral}
\end{align}
where $E$ is the detected energy, $E \in [E^\text{min}, E^\text{max}]$ corresponds to the experimental analysis window, $\Omega_{e, \nu}^\text{exp}$ is the effective solid angle for detection of electrons and (anti)neutrinos and $\mathcal{E}$ the detection efficiencies. This integration is in principle non-trivial and should ideally be performed numerically unless a high degree of symmetry exists in the experimental set-up. Additionally, since it is typically not the (anti)neutrino which is measured but instead the nuclear recoil, an additional detector response for its detection function must be introduced analogous to that of the $\beta$ particle.

For simplicity, we consider a perfect detector, i.e., $\mathcal{R}(W_e, E) = \delta(W_e-E)$ and $\mathcal{E}(\Omega) = 1$. If the experimental geometry is symmetric around, e.g., the axis of initial polarization, $\hat{J}$, we can simply perform the azimuthal integration for the $\beta$ particle and (anti)neutrino, $\int d\Omega \to 2\pi\int d \cos \theta$, leaving only the polar angle integration. Since $P_1(\cos \theta_\times)$ is odd under $\phi \to \phi + \pi$, the azimuthal integration resolves to zero.

Since at this point all further analysis depends on integration of Legendre polynomials, we introduce the following property
\begin{align}
    \int_x^1 dx^\prime P_k(x^\prime) &= \frac{1-x^2}{k(k+1)}\frac{dP_k(x)}{dx} \label{eq:legendre_int}  \\
    &\equiv I_k(x) \nonumber
\end{align}
for $k\neq 0$, so that $I_k(-1) = 0$ for all $k$, and $I_k(0)$ is $1$ for $k$ odd, and $0$ for $k$ even.

\subsubsection{Fierz cancellation in $A_\beta$}
The simplest effects can be shown in a measurement of the $\beta$-asymmetry with a single detector. Let us assume once more a perfect detector, with the ability to change the polarization direction externally. Assuming only the $\beta$ particle is detected, the integral simplifies significantly, and only $F_k^{\sigma e}$ terms remain,
\begin{equation}
    \mathcal{A}^{\uparrow(\downarrow)} = 1 + b_F \frac{1}{W_e} + \frac{1}{F_0}\sum_{k \geq 1} G_k(J_i) F_k^{\sigma e}P_k(\pm\cos \theta_e).
\end{equation}
The integrated count rates $N^{\uparrow(\downarrow)}$ are then
\begin{equation}
    N^{\uparrow(\downarrow)} = \frac{1}{2} \int_{E^\text{min}}^{E^\text{max}}dE d\Gamma_0F_0 \int_x^1 d\cos \theta_e \mathcal{A}^{\uparrow(\downarrow)},
    \label{eq:integrated_A}
\end{equation}
where $x$ denotes the polar extent of the detector. Using Eq. (\ref{eq:legendre_int}), the experimental asymmetry definition, $X$, (Eq. (\ref{eq:exp_def_X})) becomes
\begin{equation}
   X = \frac{\int dE d\Gamma_0F_0\sum_{k\text{ odd}}I_k(x)\frac{F_k^{\sigma e}}{F_0}}{\int dE d\Gamma_0F_0\left[Q + \sum_{k\text{ even}}I_k(x)\frac{F_k^{\sigma e}}{F_0} \right]},
   \label{eq:X_higher_order}
\end{equation}
where $Q = (1-x)(1+b_F/W_e)$. It is now interesting to note that since
\begin{equation}
    \frac{F_2^{\sigma e}}{F_0} \approx \frac{p_e^2}{W_e}R\nu_{12} \mathcal{S}_2 \left\{\frac{\alpha_2^0 + \alpha_2^1 W_e}{V_0^2 + A_{10}^2}\right\}
\end{equation}
using $p_e^2 = W_e^2 - 1$, additional $1/W_e$, $W_e$ and $W_e^2$ appear. The Fierz term in the denominator of Eq. (\ref{eq:X_higher_order}) consequently gets modified to
\begin{equation}
\frac{1}{W_e}\left(b_F - G_2(1+x)xR\nu_{12}\mathcal{S}_2\frac{\alpha_2^0}{V_0^2 + A_{10}^2}\right).
\end{equation}
If an integrated measurement is performed, also the effects of additional $G_2W_e$ and $G_2W_e^2$ interfere. Even in the case of a differential measurement, since $1/W_e \approx 2-W_e$ contributions from an additional $G_2 W_e$ may not be experimentally distinguishable. Assuming perfect polarization, i.e. $\langle M^2_i \rangle = J_i^2$, then $G_2 = (2-J_i^{-1})/3$. Remembering that $\alpha_2^0 \sim \mathcal{O}(10^{-2})$, cancellations on the level of $10^{-3}$ to $10^{-4}$ can occur for systems with $J_i \geq 1$. This lies in the expected sensitivity range of modern experiments.

\subsubsection{Coincidence coupling}
For most other correlations one typically measures the nuclear recoil in coincidence either with the emitted $\beta$ particle or a subsequent nuclear $\gamma$ decay. Even in the case where no energy measurement is made of the $\beta$ particle or $\gamma$, the acceptance solid angle of the secondary particle couples all other angular correlations besides the intended one, either through, e.g., the $\beta$-$\nu$ correlation or $\beta$-$\gamma$ correlation, respectively. We follow the approach by Gluck \cite{Gluck1998}.

When detecting the recoiling nucleus rather than the (anti)neutrino, we use the following identity in the center of mass frame
\begin{align}
    \int d\phi_\nu\Vec{p}_f \cdot \hat{J} &= -\int d\phi_\nu (\Vec{p}_e+\Vec{p}_\nu) \cdot \hat{J} \nonumber \\
    &= -2\pi(W_\nu \cos \theta_\nu + \beta W_e \cos \theta_e)
    \label{eq:pf_s}
\end{align}
to perform the integration of Eq. (\ref{eq:decay_rate_integration}). The latter then depends on the signs of Eq. (\ref{eq:pf_s}) and $\cos \theta_e$, leading to four different electron spectra and integrated rates. Analytical formulae for $k=1$ can be found, e.g., in Ref. \cite{Gluck1995, Gluck1998}. We can extend the results to higher orders of $k$ using the same techniques. For example, for $k=2$ the additional terms are
\begin{align}
    Q_{++}[r < 1] &= Q_{++}^1[r < 1] - \frac{f_2^{\beta\nu}}{16f_0}\left(r+r^3 \right) \nonumber \\
    &+ G_2\left\{\left(\frac{2r-r^2}{8} \right)\frac{f_2^{\sigma\nu}}{f_0} - \frac{r}{8}\frac{f_2^{\sigma e}}{f_0} \right\}, \\
    Q_{++}[r > 1] &= Q_{++}^1[r > 1] - \frac{f_2^{\beta\nu}}{16f_0}\left(1-\frac{1}{r^2}\right) \nonumber \\
    &+ G_2\left\{\frac{1}{4r}\frac{f_2^{\sigma \nu}}{f_0} +\left(\frac{1-2r^2}{8r^3}\right)\frac{f_2^{\sigma e}}{f_0} \right\},
\end{align}
where $r = p_e/W_\nu$ and $Q_{++}^1$ are, e.g., Eqs. (3.14) and (3.15) in Ref. \cite{Gluck1995} with the appropriate substitutions, and $(++)$ denotes both the $\beta$ particle and recoil going along the positive symmetry axis. We have calculated the additional terms in the infinite nuclear mass approximation, with corrections due to recoil and radiative corrections reported in Ref. \cite{Gluck1998}. The results assume perfect detection efficiency in the positive hemi-sphere, but custom results can trivially be obtained. Note that since $f_2^{\beta\nu}$ contains purely kinematical terms (see Eq. (\ref{eq:f_2_beta_nu})), it always contributes regardless of the spin change of the transition. For the neutron, however, it shows up only at the few $10^{-5}$ level \cite{Gudkov2006}.

\subsection{Real photons}
\label{sec:real_photons}
The regular $\beta$ decay process is technically always accompanied by emitted photons, so-called inner bremsstrahlung or radiative $\beta$ decay. While the branching ratio drops off steeply with increasing photon energy, the presence of the latter changes the kinematics, thereby turning $\beta$ decay into a four-body process. While this is in principle contained in the kinematic radiative corrections discussed in Sec. \ref{sec:radiative_corr}, the analysis leading to these expressions assumes the photon is either perfectly identifiable, or not detected at all. Besides measurements in calorimetric systems, a complication arises, however, when an experiment aims to measurement a correlation involving an (anti)neutrino. Taking the $\beta$-$\nu$ correlation as an example, the decay rate of Eq. (\ref{eq:general_decay_rate}) specifies a correlation for $P_k(\cos \theta_{\beta\nu})$, where
\begin{equation}
    \cos \theta_{\beta\nu} = \frac{\Vec{p_e}\cdot \Vec{p}_\nu}{|\Vec{p}_e||\Vec{p}_\nu|}.
    \label{eq:cos_neutrino}
\end{equation}
Experimentally, however, $\Vec{p}_\nu$ can typically not be measured and often one measures instead
\begin{equation}
    \cos \theta_{\beta\nu}^\text{exp} = -\frac{\Vec{p}_e\cdot (\Vec{p}_e+\Vec{p}_f)}{|\Vec{p}_e||\Vec{p}_e+\Vec{p}_f|},
    \label{eq:cos_exp}
\end{equation}
with $\vec{p}_f$ the 3-momentum of the recoiling nucleus. While these expressions are equivalent for a three-body decay, it is clear that this is not the case in the presence of an additional photon. This discrepancy was noted already a long time ago \cite{Toth1986, Gluck1990}. As a consequence, however, the formulae presented in Sec. \ref{sec:radiative_corr} involving an antineutrino are not appropriate for use in an experimental setting unless one can additionally measure the photon momentum with great accuracy. Since this is typically not a feasible option, other expressions must be derived for the radiative corrections when the emerging recoil is measured rather than the (anti)neutrino. 

A bremsstrahlung photon can arise in regular $\beta$ decay through three processes: emission from either the charged lepton, the initial or final hadronic states and from the weak vertex itself. The latter corresponds to the emission of a photon by the $W$ boson, which represents an $\mathcal{O}(G_F^4)$ process and can therefore be neglected. Due to the enormous difference in mass between the emitted charged lepton and initial and final hadronic states, practically all $\gamma$ emission arises from the charged lepton. This process is well-known to contain an infrared (IR) divergence \cite{Bloch1937}, which is cancelled by the corresponding IR divergence in the virtual photon exchange diagram \cite{Sirlin1967}, so that the two processes cannot be calculated separately. 

Due to its usefulness in experimental analyses, Gl\"uck \cite{Gluck1993} has split up the photon energy integral into a soft and hard part\footnote{Not to be confused with the separation introduced by Sirlin \cite{Sirlin1967}, which occurs at a scale between $M_A$ and $M_W$ to split low-energy QED processes from electroweak and strong physics at $M_W$.}, with an interface defined at $\omega \ll m_e$. The integration over soft photons contains the IR divergence, but the very low-energy photons do not appreciably change the kinematics, i.e. Eqs. (\ref{eq:cos_neutrino}) and (\ref{eq:cos_exp}) are quasi-identical. The radiative correction to the angular correlation term in the Dalitz distribution of Eq. (\ref{eq:dalitz}) from the virtual and soft integral can be written as \cite{Gluck1993}
\begin{equation}
    \delta a_{\beta\nu}^{VS} \approx a_{\beta\nu} \frac{\tanh^{-1} \beta}{\beta} \frac{(1-\beta^2)p_f^2}{2W_e (W_0-W_e)}.
\end{equation}

The occurrence of hard photons is responsible for a discrepancy between Eqs. (\ref{eq:cos_neutrino}) and (\ref{eq:cos_exp}), and the total decay rate can be written as \cite{Gluck1997}
\begin{align}
    \rho_H = &\frac{G_F^2}{64\pi^6}\frac{\alpha}{2\pi }\int_{1}^{W_0}dW_e\int_{\omega}^{E_\gamma}dK \nonumber \\
    &\times \int \int \int d\Omega_e d\Omega_\nu d\Omega_\gamma K\beta W_\nu W_e |M_{\gamma}|^2,
    \label{eq:hard_photon}
\end{align}
where the matrix element can be written as
\begin{equation}
    |M_\gamma|^2 = f_0^{LO} \biggl[ H_0(K^\mu) + a_{\beta\nu}H_1(K^\mu) \biggr]
    \label{eq:M_gamma_decomp}
\end{equation}
with $K^\mu$ the photon four-momentum, $f_0^{LO}$ the leading-order isotropic spectral function (Eq. (\ref{eq:f_0_LO})) and the expressions for $H_i$ can be found, e.g., in Refs. \cite{Gluck1993, Gluck1997}. More specifically, $H_0(K^\mu)$ is a simple scalar and
\begin{subequations}
\begin{align}
    H_0(K^\mu) &= \mathcal{A}, \label{eq:H_0_MBR}\\ 
    H_1(K^\mu) &= \vec{p}_e \cdot \vec{p}_\nu\, \mathcal{B} + \vec{p}_\nu \cdot \vec{K}\, \mathcal{C},
    \label{eq:H_1_MBR}
\end{align}
\end{subequations}
where $\mathcal{A}, \mathcal{B}$ and $\mathcal{C}$ are scalars depending on kinematics, and $\mathcal{A} \sim \mathcal{B} \gg \mathcal{C}$.

From an experimental point of view, it is most interesting to note that both $H_0$ and $H_1$ contain collinear peaks due to the charged particle propagators. Since the $\beta$ particle is mainly responsible for the emission of real photons, it implies that the photon distribution is peaked along the $\beta$ particle direction. For $1 \ll W_e$ and $\theta_{\beta\gamma} \ll 1$ the matrix element has the following behavior \cite{Gluck1997}
\begin{equation}
    K|M_\gamma|^2 \sim \frac{1}{\theta_{\beta\gamma}^2+W_e^{-2}},
    \label{eq:M_gamma_collinear}
\end{equation}
where $\theta_{\beta\gamma}$ is the angle between the photon and outgoing lepton in the center of mass frame\footnote{While for regular $\beta$ decay $1 \ll W_e$ is often not valid due to the low energy transfer, the peaking of the photon distribution along the $\beta$ momentum arises naturally from Lorentz invariance.}.

\begin{figure}[h!]
    \centering
    \includegraphics[width = 0.48\textwidth]{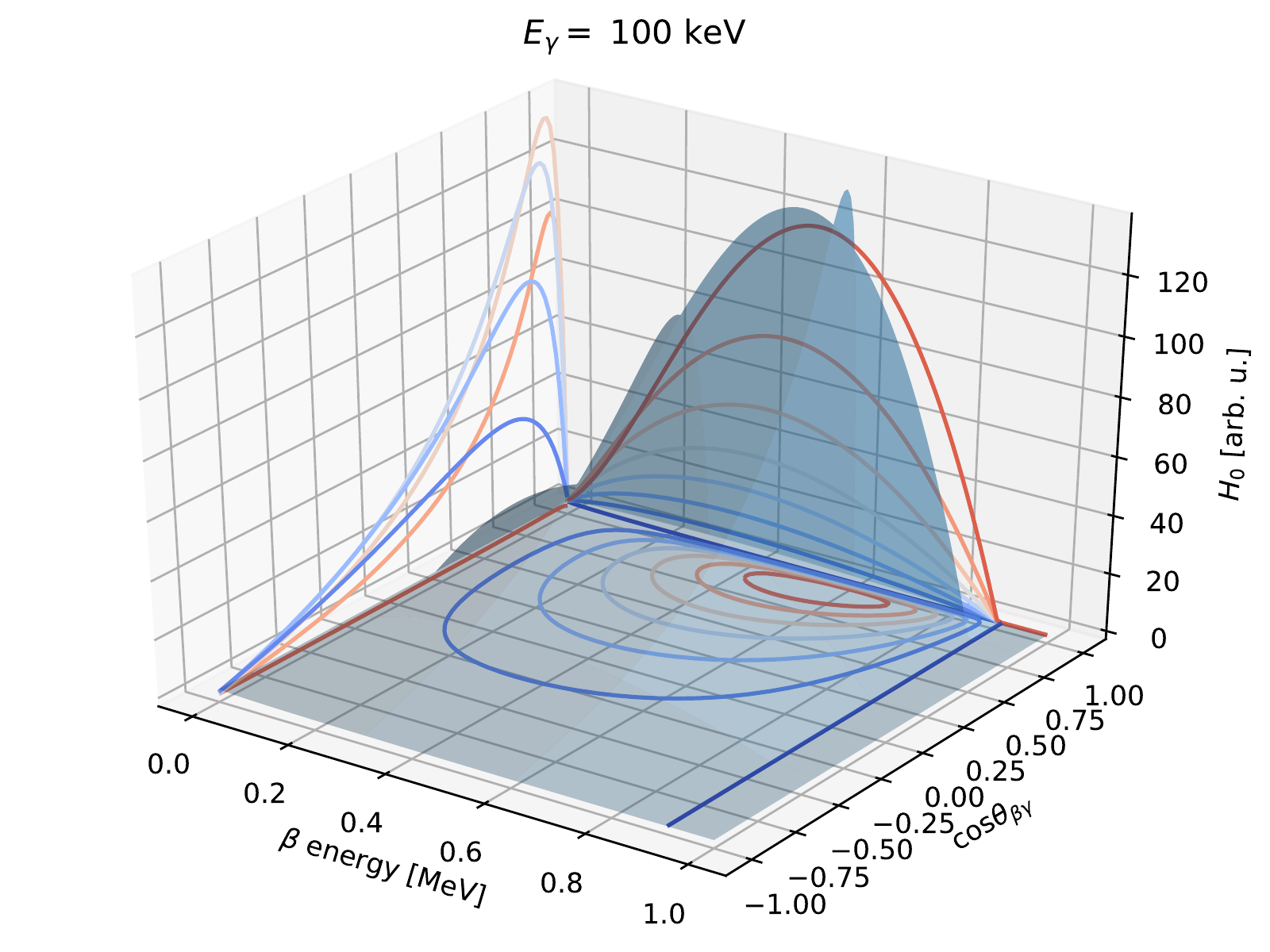}
    \caption{Behaviour of $H_0$ from Eq. (\ref{eq:M_gamma_decomp}) for a fixed photon energy of $100$ keV in a 1 MeV $\beta$-transition, as a function of $\beta$ energy and $\cos \theta_{\beta\gamma}$. }
    \label{fig:Hard_BR_H0}
\end{figure}

Figure \ref{fig:Hard_BR_H0} shows the behavior of $H_0$ as a function of $\beta$-energy and $\cos \theta_{\beta \gamma}$. As expected, a maximum is reached for nearly-collinear $\beta$ particle and photon, with an approximate parabolic behaviour near $\cos \theta_{\beta\gamma} = 1$ as in Eq. (\ref{eq:M_gamma_collinear}). Due to the addition of two extra degrees of freedom ($\cos \theta_{\beta \nu}$ and $\cos \theta_{\nu \gamma}$), $H_1$ is not as straightforward to show graphically. The kinematic structure is very similar to $H_0$, however, as was shown in Sec. \ref{sec:radiative_corr}.

Unfortunately, however, analytical solutions to Eq. (\ref{eq:hard_photon}) are typically not available when one detects the recoiling nucleus. That is because the analytical results obtained by several authors \cite{Yokoo1973, Yokoi1976, Morita1976, Garcia1982} calculate the inner bremsstrahlung corrections for a constant (anti)neutrino energy and treat the final nucleus as infinitely massive, thereby simplifying Eq. (\ref{eq:hard_photon}) substantially. This corresponds to integrating over the photon energy and direction keeping $\vec{p}_e$ and $\vec{p}_\nu$ constant, making Eqs. (\ref{eq:H_0_MBR}) and (\ref{eq:H_1_MBR}) straightforward. Those results give rise to Eq. (\ref{eq:decay_rate_RC}), with the small difference in $\mathcal{A}$ and $\mathcal{B}$ resulting in the near-equality of $g(W, W_0)$ and $h(W, W_0)$ in Eqs. (\ref{eq:g_sirlin}) and (\ref{eq:h_shann}) (see also Fig. \ref{fig:RC}). While semianalytical results have been reported \cite{Gluck1993}, those assume perfect reconstruction of $\beta$ particle and recoil energies in a closed $4\pi$ geometry with perfect detectors. The only way then to take into account experimental conditions is through a numerical procedure, such as that outlined in Ref. \cite{Gluck1997, Gluck1997a}, or through explicit event generation of the additional photon according to Eq. (\ref{eq:M_gamma_decomp}) \cite{HayenCRADLE}.

While the above discussion focused on a measurement of the $\beta$-$\nu$ correlation, the same argument applies to the $\nu$-asymmetry, $B_\nu$ and the $\beta$-$f$ correlation. Some published results are available for the neutron \cite{Gluck1992, Gluck1990} and recoil spectra of $^6$He and $^{32}$Ar \cite{Gluck1998a}. Further analysis is planned for future work.


\subsection{Effective coefficient}
\label{sec:effective_coef}
Equation (\ref{eq:coeff_X}) introduced an effective correlation coefficient, $\widetilde{X}$, as the way many experiments analyse the sensitivity to exotic scalar or tensor currents through the Fierz term. When the correlation involves the (anti)neutrino, however, some caveats once again emerge as one typically measures the nuclear recoil rather than the antineutrino. The case that was extensively described in Ref. \cite{Gonzalez-Alonso2016} discusses the measurement of the $\beta$-$\nu$ asymmetry through the measurement of the recoil energy distribution (see Eq. (\ref{eq:dalitz})). Due to the dependence of the $\beta$ energy on the $\beta$-$\nu$ angle (Eq. (\ref{eq:cos_beta_nu_recoil})), a measurement of only the latter does not allow for a parametrization of Fierz in terms of $\widetilde{X}$. More generally, Eq. (\ref{eq:coeff_X}) is only valid when a correlation described as $XR(W_e, \theta)$ with $R$ any function, the observables $W_e$ and $\theta$ are separable, i.e., when measuring only the angular variable
\begin{align}
    \frac{d\Gamma}{d\theta} &= \int dW_e G(\theta)H(W_e)\left[1 + b_F \frac{1}{W_e} + XR(W_e, \theta) \right] \nonumber \\
    &= C G(\theta)\left(1 + b_F\langle\frac{1}{W_e}\rangle \right)\left[1 + \widetilde{X}\langle R(W_e, \theta)\rangle_{W_e} \right].
\end{align}
When $a_{\beta\nu}$ is determined from the recoil momentum only, this is not the case as $\theta$ and $W_e$ are coupled through momentum conservation (Eq. (\ref{eq:cos_beta_nu_recoil})). In that case, a parametrization like $\widetilde{X}$ is not valid, and one has to properly integrate over the Dalitz distribution of Eq. (\ref{eq:dalitz}) according to the experimental geometry to obtain the correct result. Alternatively, the $\beta$ energy is determined and a fit is performed either on the Dalitz distribution, or on slices of constant $\beta$ energy \cite{Broussard2017a, Broussard2019}.

\subsection{Practical difficulties}
In any real experiment there's a potentially large number of additional practical difficulties, such as detector non-linearities, detection efficiencies, energy losses outside of the active detector area, unresolved polarization and/or alignment, etc. Regardless of the conceptual problems posed above, all of these must be overcome in order to extract meaningful results. Clearly, the precise occurrence of each of these effects is unique to each experiment. In this section we describe two critical contributions to asymmetry measurements, and demonstrate their particular advantage compared to, e.g., spectrum measurements.

As an example, consider an experiment in which the calibrated energy defines the boundaries of energy bins, with a $\beta$-correlation calculated in each bin according to Eq. (\ref{eq:exp_def_X}). In the case of any physical detector, there will generally be some remaining systematic difference between the deposited energy $E$ and the reconstructed energy $\epsilon$, such that
\begin{equation}
    E = \epsilon + \sum_{i = 0}c_i \epsilon^i,
    \label{eq:E_calibration}
\end{equation}
where $c_i$ are parameters that can be constrained based on the calibration procedure in place. Clearly, a larger number of calibration points and good representation throughout the region of interest will force the different $c_i$ to be smaller. In the following we assume this uncertainty to be sufficiently small, i.e. $|(\epsilon-E)/E| \ll 1$, and neglect losses causing $E$ to be smaller than the initial energy. We assume a general decay rate
\begin{equation}
    \frac{d\Gamma}{dE} = K(E, \theta)\left[1+\frac{b_F}{E}+\chi(E, \theta)\right],
\end{equation}
where $\theta$ is some angle and $\chi$ switches sign for ``up" and ``down" detectors. The measured bin counts are then
\begin{equation}
    N^\uparrow = \Delta t\int d\Omega^\uparrow \int_{f(\epsilon_l)}^{f(\epsilon_h)} dE K(E, \theta)\left[1 + \frac{b_F}{E} + \chi(E, \theta) \right]
    \label{eq:N_up_calibration}
\end{equation}
where $\Delta t$ is the measurement time and $f$ is Eq. (\ref{eq:E_calibration}) for the low and high bin edges $\epsilon^l, \epsilon^h$, respectively. The $\beta$ correlation after performing the angular integration is then
\begin{equation}
  X(\bar\epsilon) = \frac{N^\uparrow - N^\downarrow}{N^\uparrow + N^\downarrow} = \frac{\int ^{E^h } _{E^l } \langle K(E)\chi(E) \rangle_{\Omega}  dE }{\int ^{E^h } _{E^l } \langle K(E) \rangle_\Omega (1+b_F/E) dE },
\end{equation}
where $\bar\epsilon$ denotes the bin center, and $E^{h, l} \equiv f(\epsilon^{h, l})$. We introduce some additional notational simplicity
\begin{subequations}
\begin{align}
    \mathcal{A}(E) &= \langle K(E)\chi(E) \rangle_\Omega \label{eq:A_asym} \\
    \mathcal{S}(E) &= \langle K(E) \rangle_\Omega (1+ b_F/E)
    \label{eq:S_sym}
\end{align}
\end{subequations}
to denote the (a)symmetric (numerator) denominator integrand. We then perform a Taylor expansion of $\mathcal{A}, \mathcal{S}$ around $E^l$ so that
\begin{equation}
    X(\bar \epsilon) = \frac{\sum_{n} \mathcal{A}^{(n)}(E^l)(\Delta E)^{n+1}/(n+1)!}{\sum_n \mathcal{S}^{(n)}(E^l)(\Delta E)^{n+1}/(n+1)!}
\end{equation}
with 
\begin{equation}
    \Delta E \equiv E^h-E^l = \epsilon^h-\epsilon^l + \sum_{i=1}c_i\left[ (E^h)^i-(E^l)^i\right]
\end{equation}
and superscript $(n)$ denoting the $n$-th derivative. The systematic difference induced by the possible calibration errors can be then written as
\begin{equation}
    \Delta X = \frac{\sum_{n, m}\frac{\mathcal{A}^{(n)}(E^l) \mathcal{S}^{(m)}(\epsilon^l) (\Delta E)^{n+1}(\Delta \epsilon)^{m+1} - E\leftrightarrow \epsilon}{(n+1)!(m+1)!}}{\sum_{n, m}\frac{\mathcal{S}^{(n)}(E^l) \mathcal{S}^{(m)}(\epsilon^l) (\Delta E)^{n+1} (\Delta \epsilon)^{m+1}}{(n+1)!(m+1)!}}
    \label{eq:delta_X_calibration}
\end{equation}
with $\Delta \epsilon = \epsilon^h - \epsilon^l$. Equation (\ref{eq:delta_X_calibration}) is not terribly enlightening, so it is worthwhile to consider some examples. 

In the case where $\chi$ is energy-independent, is it obvious from Eqs. (\ref{eq:A_asym}) and (\ref{eq:S_sym}) that the numerator only survives for non-zero $b_F$. As a consequence, the observable effect is $\mathcal{O}\left(b_F \{\Delta E-\Delta\epsilon\}\right)$ and can be neglected. This is in stark contrast to when one measures only the spectrum for a Fierz extraction, where a handle on Eq. (\ref{eq:E_calibration}) is crucial \cite{Hickerson2017}. Taking the $\beta$-asymmetry as another example, we have up to leading order 
\begin{equation}
    \chi(E, \theta) \approx \beta A^{LO}_\beta \langle P \rangle \cos \theta_{e},
\end{equation}
with $A_\beta^{LO}$ as in Eq. (\ref{eq:A_LO}) and $\langle P \rangle$ the average polarization. The energy-dependence at this order comes only from $\beta$, so that the effects of calibration uncertainty are mainly relevant at lower energies, since $\beta^{(1)} \to 0$ as $\beta \to 1$. Going beyond leading order, additional energy dependence shows up coming from induced currents (Eq. (\ref{eq:f_sigma_e})) and radiative corrections (Eq. (\ref{eq:R_approx})). As these are themselves small corrections of $\mathcal{O}(10^{-3})$, assuming the calibration to be sufficiently under control, these can again be neglected. 

The situation changes when one takes into account detection efficiencies. This efficiency, typically determined by scattering effects and the detector threshold, can differ significantly from unity, particularly near-threshold. This fact is exacerbated by the typically strong angular dependence of the efficiency due to, e.g., dead layer losses and backscattering. As a consequence the angular integration in Eq. (\ref{eq:N_up_calibration}) should be replaced 
\begin{equation}
    \int d\Omega \int dE \to \int \int d\Omega\, dE~\mathcal{E}(E, \Omega) 
\end{equation}
leading to corresponding changes in Eqs. (\ref{eq:A_asym}) and (\ref{eq:S_sym}). Due to the different angular weighting of the (a)symmetric distributions, efficiency differences show up to first order when using $X$ as in Eq. (\ref{eq:exp_def_X}). If one formulates the measurement in terms of a ratio of rates for two different spin states for a single detector or a "super-ratio" of two or more detectors arranged with the appropriate spin dependence, however, one expects first order cancellation for constant efficiency factors shared by both isotropic and $\cos \theta$-weighted distributions. Relative corrections for the isotropic to angular distributions (due to, e.g., backscattering effects) do not cancel, however.  As with calibration errors, these appear in the measured asymmetry at the level of the difference in these efficiencies and energy dependent effects for isotropic {\it vs.} $\cos \theta$-weighted decays, and the differences in the integrals over energy (which vanish as $\beta \rightarrow 1$).

Given that all effects for calibration errors and efficiency differences smoothly scale with the energy, these also scale with the bin-size. As a consequence, effects can be minimized by preparing super-ratios with the smallest reasonable bin-sizes. The generic suppression of sensitivity to calibration errors and detection efficiencies in asymmetry measurements are significant advantages for their use in placing direct limits on BSM contributions to $\beta$ decay\footnote{This advantage provided significant motivation for early work on the PERKEO experiment \cite{Bopp1986, Bopp1988} and the 19Ne $\beta$ asymmetry measurement \cite{Calaprice1975a}.}.

\section{New Physics Sensitivity}
\label{sec:new_physics_sensitivity}

As already mentioned in the introduction, precision measurements of correlation coefficients in (nuclear) $\beta$ decay are an attractive option for new physics searches both from a theory and experimental point of view due to their relative nature. The latter allows for a cancellation of many systematic uncertainties in an experimental setting, and theoretically it is often easier to reliably estimate ratios of matrix elements than their absolute magnitude. As a consequence, the reach for these measurements can be fairly broad. We focus here on two possible cases, namely the search for exotic currents through the appearance of a Fierz term, and $V_{ud}$ determinations for certain mirror systems and CKM unitarity.

\subsection{$V_{ud}$ and CKM unitarity}
Common to all semileptonic $\beta$ decays, the decay rate is determined at the coupling level by the following product
\begin{equation}
    \Gamma_{semi-l} \propto G_F^2\, V_{ud}^2~ g_V^2(1+ \Delta_R) F
\end{equation}
where $G_F$ is the Fermi coupling constant, $V_{ud}$ is the $up$-$down$ quark mixing matrix element, $g_V^2(1+\Delta_R)$ is the renormalized vector coupling constant and where $F$ takes into account additional transition-specific information. If all other information can be either experimentally or theoretically sufficiently determined, a measurement of the lifetime in semileptonic systems gives access to $|V_{ud}|$. The $\beta$-decay of the muon is theoretically extremely clean, which allows one to calculate both $F^\mu$ and $\Delta_R^\mu$ very accurately. The latter is lumped together with $G_F$ to define the traditional Fermi coupling constant which is experimentally found to be $G_F = 1.1663788(7) \times 10^{-5}$ GeV$^{-2}$ \cite{Webber2011}. In neutron or nuclear systems then, the conserved vector current hypothesis sets $g_V = 1$ up to higher-order corrections \cite{Ademollo1964}, and one needs to calculate only that part of $\Delta_R$ which is unique to nuclei \cite{Sirlin1978, Sirlin2013}. Finally, in the nuclear sector $F$ can be calculated to high precision in two different cases: ($i$) superallowed $0^+ \to 0^+$ Fermi decays \cite{Hardy2015}, and ($ii$) $T=1/2$ mirror decays \cite{Severijns2008, Naviliat-Cuncic2009}. In both cases the Fermi matrix element is dictated by isospin symmetry and one finds $M_F
^0$ equal to $\sqrt{2}$ and $1$, respectively. The experimental input needed for $F$ in both cases consists of the half-life of the $\beta$ transition, the branching ratio, and the endpoint energy. Because mirror decays have both non-zero Fermi and Gamow-Teller components and the latter is not constrained by symmetry, this needs to additionally be experimentally determined from, e.g., the measurement of a $\beta$-correlation (see Sec. \ref{sec:mirror}).

One can construct a so-called $\mathcal{F}t_0$ value, analogous to the $\mathcal{F}t$ for superallowed decays \cite{Hardy2015}, which according to CVC must be equal for all nuclear mirrors \cite{Severijns2008}
\begin{align}
    \mathcal{F}t_0 &\equiv f_Vt(1+\delta_R)(1+\delta_{NS}-\delta_C)\left(1+\rho^2 \frac{f_A}{f_V}\right) \nonumber \\
    &= \frac{K}{g_V^2G_F^2V_{ud}^2}\frac{1}{|\mathcal{M}_F^0|^2(1+\Delta_R^V)}
    \label{eq:Ft0}
\end{align}
where $K = 8120.278(4) \times 10^{-10}$ GeV$^{-4}$ s is a combination of constants, $\delta_i$ correspond to radiative ($R$), isospin-breaking ($C$) and nuclear structure ($NS$) corrections, and $f_{V, A}$ are so-called phase space integrals \cite{Towner2015, Hayen2018, Hayen2019c}. The first line shows all transition-specific factors, while the second line consists only of common constants.

Because of strong cancellations in some mirror transitions such as the neutron and $^{19}$Ne, great sensitivity can be obtained for a determination of $\rho$ from a $\beta$ correlation coefficient. Following a reduction in uncertainty and removal of double counting in $f_A/f_V$ calculations \cite{Hayen2019c}, precise measurements of mirror transitions can shed light both on the shift in inner radiative corrections throughout the lower mass region, and be competitive in setting constraints on CKM unitarity, 
\begin{equation}
    \Delta_{CKM} = |V_{ud}|^2 + |V_{us}|^2 + |V_{ub}|^2 - 1.
    \label{eq:delta_CKM}
\end{equation}

\begin{figure}
    \centering
    \includegraphics[width=0.48\textwidth]{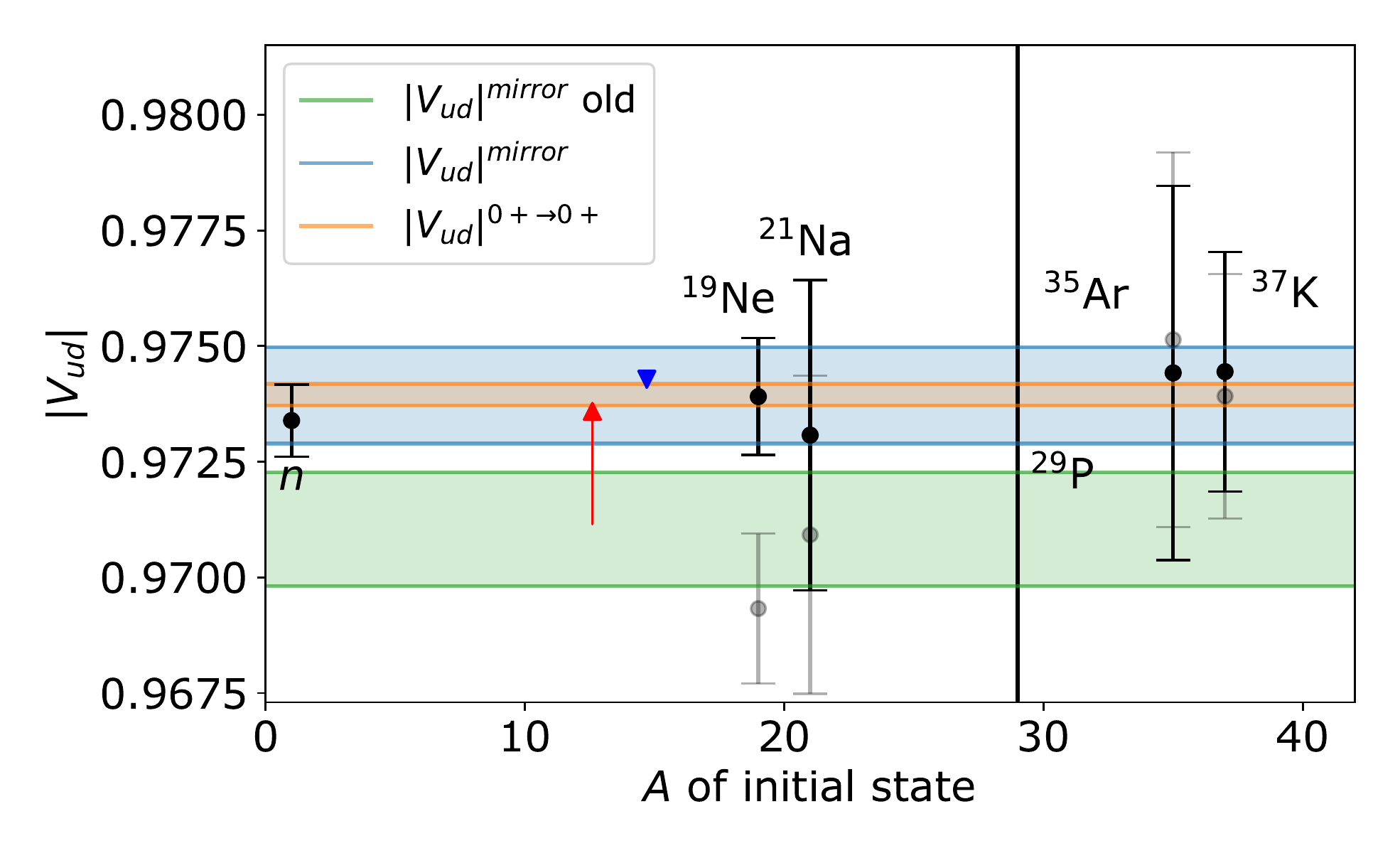}
    \caption{Summary of current status and the influence of theory changes on the mirror $|V_{ud}|$ extraction compared to the superallowed $0^+\to 0^+$ data set and the neutron. The blue arrow signifies the shift in $|V_{ud}|^\text{mirror}$ due to the change in $\Delta_R^V$ \cite{Seng2019b}, while the red arrow signifies the shift due to updated $f_A/f_V$ values \cite{Hayen2019c}, which is also shown in grey for individual results. The uncertainty in $^{19}$Ne is significantly smaller because of the reduced theory uncertainty from $f_A/f_V$ results \cite{Hayen2019c}.}
    \label{fig:Vud_mirrors}
\end{figure}

Figure \ref{fig:Vud_mirrors} shows the current status and summary of changes in the past years for the $|V_{ud}|$ extraction from nuclear mirrors and its comparison to the neutron and superallowed decays \cite{Hardy2015, Seng2019b}. Recently, renewed attention has been devoted to the calculation of the inner radiative correction \cite{Seng2018, Gorchtein2018}, with a significant shift in its central value since its last evaluation in 2006 \cite{Marciano2006}. While this shift is consequential for the superallowed decays and to an extent the neutron, the results of the mirror decays are most significantly impacted by change in $f_A/f_V$ values \cite{Hayen2019c, HayenGTRC}. Because of the reduced uncertainty on the latter, the new experimental measurement of $\rho$ for $^{19}$Ne \cite{Combs2020} is significantly lower and dominated by experiment \cite{HayenGTRC}.

In addition to unitarity tests, the required internal consistency of the $\mathcal{F}t_0$ values in mirror and super-allowed decays provides a number of paths to very clean constraints for new physics, some of which can evade the precision limits imposed by the vertex corrections. Examples include the Marciano's axial coupling relationship for the neutron \cite{Czarnecki2018}, a neutron lifetime consistency test \cite{Pattie2013, *Pattie2015}, the ratio of $|V_{ud}|$ values extracted from the neutron and the superallowed decays articulated by \cite{Bhattacharya2012}.

\subsection{Exotic currents}
Traditionally the search for exotic currents in low energy nuclear $\beta$ experiments have been interpreted in terms of the Lee-Yang Hamiltonian \cite{Lee1956}. The past decade has seen tremendous progress in the development of effective field theories at the quark level, which allows one to directly compare obtained limits to LHC constraints if the new physics lies above the LHC energy scale. Neglecting right-handed neutrinos and writing only linear BSM couplings, one can write \cite{Gonzalez-Alonso2018, Cirigliano2013a}

\begin{align}
    \mathcal{L}_\text{eff} &= - \frac{G_F\tilde{V}_{ud}}{\sqrt{2}}\biggl\{\bar{e}\gamma_\mu(1-\gamma^5)\nu_e \cdot \bar{u} \gamma^\mu[1-(1-2\epsilon_R)\gamma^5]d \nonumber \\
    &+ \epsilon_S\, \bar{e}(1-\gamma^5)\nu_e \cdot \bar{u}d - \epsilon_P\, \bar{e}(1-\gamma^5)\nu_e \cdot\bar{u}\gamma^5 d \nonumber \\
    &+ \epsilon_T\, \bar{e}\sigma_{\mu\nu}(1-\gamma^5)\nu_e \cdot \bar{u} \sigma^{\mu\nu}(1-\gamma^5)d \biggr\} + \text{h.c.},
    \label{eq:L_eff}
\end{align}
with
\begin{equation}
    \tilde{V}_{ud} \approx V_{ud}\left(1+\epsilon_R + \epsilon_L - \frac{\delta G_F}{G_F}\right) \label{eq:Vud_tilde}
\end{equation}
and $\epsilon_i$ are linear BSM effects of order $\mathcal{O}(M_W^2/\Lambda_{BSM}^2)$ and $\delta G_F$ contains new physics contributions specific to muon decay. Equation (\ref{eq:Vud_tilde}) is what causes a deviation from CKM unitarity at the quark level, i.e. Eq. (\ref{eq:delta_CKM}), which can, e.g., be investigated using the $\mathcal{F}t_0$ values for mirror and superallowed decays.

From Eq. (\ref{eq:L_eff}) one can see that the axial coupling constant is renormalized at the quark level, so that a measurement of $\rho$ in different systems cannot reveal some BSM physics, as one always measures $\tilde{\rho} = \rho(1-2\epsilon_R)$. An exception to this is when the coupling constant can be calculated to high precision from theory, which is only feasible for the neutron using lattice QCD (LQCD). The determination of $\tilde{g}_A$ from a measurement of $\tilde{\rho} = \sqrt{3}\tilde{g}_A/g_V$ in the neutron is theoretically a clean channel for looking for right-handed currents through the comparison with LQCD \cite{Chang2018, Gonzalez-Alonso2018}
\begin{equation}
    \tilde{g}_A = g_A^{LQCD}[1-2\, \text{Re}(\epsilon_R)],
    \label{eq:gA_LQCD}
\end{equation}
where care must be taken to take into account the difference in inner radiative corrections between vector and axial vector parts \cite{Hayen2019c}. The constraints from Eq. (\ref{eq:gA_LQCD}) are currently limited by the uncertainty on LQCD results, which vary between 1\% and 4\% \cite{Chang2018, Gupta2018}.

We note in passing that while both the older \cite{Calaprice1975a} and newer \cite{Combs2020} measurements of $A_\beta$ in $^{19}$Ne allow for a non-zero induced tensor component - a so-called Second-Class Current (SCC), see App. \ref{app:coefficients} - through a two-parameter fit of the slope in $A_\beta$, their findings are of opposite sign and the modern result is not statistically significant ($1\sigma$). An SCC would additionally show up as a difference in $\lambda$ extracted from $a_{\beta\nu}$ and $A_\beta$ in the neutron \cite{Gardner2001}. While there is currently some tension between the results of PERKEO III \cite{Markisch2019} and aSPECT \cite{Beck2019}, a simplified analysis shows that SCC effects would need to be about three times larger than that expected from weak magnetism and result in a significantly higher value of $\lambda \approx 1.288$ for both to agree. The latter would be in strong violation of CKM unitarity and the additional data sets as in, e.g., Fig. \ref{fig:Vud_mirrors} and point to the presence of right-handed currents from Eq. (\ref{eq:gA_LQCD}) when using \cite{Chang2018} at face value. We conclude that at this time there is no strong evidence for SCCs.

Finally, we show a simple analysis demonstrating the physics reach of mirror decays in the search for scalar and tensor currents. Several experimental programs are currently underway to measure or constrain $b_F$ to a few parts in $10^3$ using either the $\beta$-asymmetry or $\beta$-$\nu$ correlation, predominantly in the neutron \cite{Gonzalez-Alonso2018}. Here we make use of the fact that the Fierz term changes sign for $\beta^\pm$ decay as in Eq. (\ref{eq:bF}) to compare the $\mathcal{F}t_0$ values of the neutron and $^{19}$Ne and obtain competitive limits.

By turning on BSM physics, Eq. (\ref{eq:Ft0}) is modified according to
  \begin{align}
    \mathcal{F}t_0 &= f_V t \left( 1 + \delta _r ' \right ) \left(1 + \delta _{NS}^V - \delta _{C}^V \right ) \left(1 + \frac{f_A}{f_V}{ \tilde{\rho}} ^2 \right) \nonumber \\
    &=  \frac{K}{g_V ^2 G_F^2\vert V_{ud} M_F ^0 \vert ^2 \left ( 1 + \Delta _{R} ^V \right )} \nonumber \\
    & \times  \frac{1}{\left[ 1 + 2\epsilon _L + 2\epsilon _R - 2\frac{\delta G_F}{G_F} + b_F \langle W^{-1} \rangle \right]},
    \label{eq:Ft0_BSM}
    \end{align}
where all BSM physics is contained in the last line and the Fierz contribution is transition-dependent due to the endpoint-dependence on $\langle W^{-1} \rangle$. As a consequence, a ratio of $\mathcal{F}t_0$ values for the neutron and $^{19}$Ne maintains sensitivity only to the Fierz term, but all other common theoretical inputs cancel. The change in sign in $b_F$ enhances its sensitivity. 

While Eq. (\ref{eq:Ft0_BSM}) is correct, there is an additional subtlety involved when using experimental input for $\tilde{\rho}$. As an example, we discuss its extraction from the $\beta$-asymmetry. As mentioned in Eq. (\ref{eq:coeff_X}), the presence of a non-zero Fierz term serves to dilute experimentally observed $\beta$-asymmetry
\begin{equation}
  A_\beta^\text{exp} = \frac{A^{SM}_\beta}{1 + b_F \langle W^{-1} \rangle^\text{exp}}
\end{equation}
where the ``exp" superscript serves as a reminder that the average is calculated over the experimentally analysed range rather than the full spectrum. 
Our measured ratio of $\mathcal{F}t_0$ values must therefore be modified (where we use the subscript ``m'' now for measured values):
\begin{align}
    \mathcal{F}t_{0,m} &\equiv \mathcal{F}t_0\left[ \frac{1 + \frac{f_A}{f_V}{\tilde \rho_m} ^2}{1 + \frac{f_A}{f_V}{\tilde \rho} ^2} \right ] \nonumber \\
    &\approx \mathcal{F}t_0\left[1 + 2\frac{\tilde{\rho}}{1+\tilde{\rho}^2} \frac{d\tilde{\rho}}{db_F}b_F \right ] \nonumber \\
    &=  \frac{K}{g_V^2 G_F^2 \vert V_{ud} M_F ^0 \vert ^2 \left ( 1 + \Delta _{R} ^V \right )} \nonumber \\
    & \times \frac{\left [1 - 2\dfrac{\tilde{\rho}}{1+\tilde{\rho}^2} A_\beta b_F \langle W^{-1} \rangle^\text{exp} \dfrac{d\tilde{\rho}}{dA} \right ]}{\left [ 1 + 2\epsilon _L + 2\epsilon _R - 2\frac{\delta G_F}{G_F} + b_F \langle W^{-1} \rangle \right ]},
\end{align}
where we take $f_A/f_V \approx 1$ and up to linear order in BSM couplings $\tilde{\rho} b_F = \tilde{\rho}_m b_F$. The ratio of $\mathcal{F}t_0$ values for $^{19}$Ne and the neutron can then be written as
\begin{align}
  \mathcal{R}_m &\equiv \frac{\mathcal{F}t_{0,^{19}\text{Ne}}}{\mathcal{F}t_{0,n}} \nonumber \\
  &=\frac{1 +  b_F^n\left[\langle W^{-1}\rangle  + 2\dfrac{\tilde{\rho}}{1+\tilde{\rho}^2}\dfrac{d\tilde{\rho}}{dA}A_\beta^\text{SM} \langle W^{-1} \rangle^\text{exp}\right ]_n}{1 + b_F^{19}\left[ \langle W^{-1} \rangle + 2\dfrac{\tilde{\rho}}{1+\tilde{\rho}^2} \dfrac{d\tilde{\rho}}{dA}A_\beta^\text{SM} \langle W^{-1} \rangle^\text{exp} \right ]_{^{19}\text{Ne}}}.
\end{align}

Using the recent measurements of $A_\beta$ for the neutron \cite{Brown2018, Mund2013, Markisch2019} and $^{19}$Ne \cite{Combs2020} as numerical input and evaluating $b_F$ using Eq. (\ref{eq:bF}) with the latest lattice charges, we find\footnote{We evaluate $b_F$ at $\tilde{\rho}_m$ and neglect the details of the energy dependence of the Fierz term in the one parameter fit to the asymmetry.  These introduce a systematic error in our extracted value of $\epsilon _T$ less than about 4\% for the neutron and $^{19}$Ne for $|b_F| < 0.01$.}
\begin{align}
  \mathcal{R}_m &= \frac{1 + (-5.2\epsilon_T + 0.33\epsilon_S)(0.65 + 0.23)}{1  - (-4.5\epsilon_T + 0.54\epsilon_S)(0.39 - 0.04)} \nonumber \\
  &\approx 1 - 6.1\epsilon_T + 0.49\epsilon_S.
\end{align}

If we use values for $^{19}$Ne of $\mathcal{F}t_{0,19} = 6142(17)$ and the neutron of $\mathcal{F}t_{0,n} = 6155.6(65)$ \cite{Combs2020}, their ratio is $\mathcal{R}=0.9978(29)$, where the uncertainty is dominated by that of the $^{19}$Ne $\mathcal{F}t_0$ value, which it itself dominated by that on $\tilde{\rho}$. Using a value of $\epsilon_S = 1.4(2.0)\times 10^{-3}$ from Hardy and Towner \cite{Hardy2015}, we determine $\epsilon_T = 4.8(7.3)\times 10^{-4}$ (90\% C.L.).  The resulting limit for the BSM energy scale for an exotic tensor coupling at the 90\% (C.L.) is $\Lambda _T > 5.1$~TeV. Using instead the full neutron decay data set, we get $\mathcal{F}t_0 = 6151(10)$, which yields a ratio $\mathcal{R} = 0.9990(31)$. This ratio agrees with the value determined using beta asymmetry measurements only and reduces the lower limit of the energy scale by 5\% due to the increased uncertainty in $\tilde{\rho}_{m,n}$. If the uncertainty due to $^{19}$Ne is brought to the same level as that of the neutron (an improvement of a factor $2.5$) through an improved measurement of $\tilde{\rho}$, the tensor scale becomes $\Lambda_T > 7.2$\,TeV.

\section{Conclusion}
\label{sec:conclusion}
Measurements of correlations in (nuclear) $\beta$ decay have continuously been a central pillar in the exploration of the low energy electroweak sector of the Standard Model, and modern experiments are entering a regime where additional theory corrections become relevant. We have compiled here a comprehensive summary of theory input with a special focus on the $\beta$-asymmetry ($A_\beta$) and $\beta$-$\nu$ correlation ($a_{\beta\nu}$). In particular, we have reviewed the kinematic and nuclear structure effects, including those of higher order in the relevant angle and electroweak radiative corrections. We have taken another look at the mirror $T=1/2$ systems for their experimental interest and theoretical simplifications. 

A significant portion of this work discussed effects specific to an experimental setting, in particular because many higher order effects at the limit of current experimental precision are dependent upon the experimental geometry and detection mechanisms. Depending on the system, neglecting one or more of these effects can, e.g., mask or reduce sensitivity to a Fierz term, and skew results away from Standard Model expectations even in the absence of new physics. We have provided a number of examples, both conceptual and practical, of when these occur in a simplified detector description along with more general results, so that they may be understood and employed in more complex geometries and detection schemes. 

Finally, we have shown how the input from $\beta$-correlation measurements feeds into tests for new physics in the electroweak sector. This was done with a particular focus on mirror decays, which have a number of pleasant features which make them prime candidates for high-impact measurements. Together with the neutron, these have undergone steady progress over the last decade and have the potential to become as precise as the superallowed data set, with different systematic uncertainties. Additionally, we have shown in a simplified analysis that by using only the neutron and $^{19}$Ne a sensitivity on new tensor couplings lies above $5.8$\,TeV (90\% C.L.), with the potential to lift it above $8$\,TeV through an improvement of only the mixing ratio for $^{19}$Ne. Given the subtleties at this level of scrutiny, it is to the benefit of experiments to use a comprehensive formalism.

\begin{acknowledgements}
The authors would like to acknowledge useful discussion with and and inspiration from V. Cirigliano, A. Garcia, O. Naviliat-Cuncic, B. M\"arkisch, W. Marciano, D. Melconian, B. Plaster, G. Ron, and N. Severijns. This article was supported through the Department of Energy, Low Energy Physics grant DE-FG02-ER41042 and NSF grant PHY-1914133.
\end{acknowledgements}

\appendix

\section{Notation and conventions}
\label{app:notation_conventions}
In the entirety of the manuscript we use units suitable to $\beta$ decay, i.e.
\begin{equation}
    \hbar = c = m_e = 1.
\end{equation}
As a consequence, typical $\beta$ energies are of order unity, the nuclear radius $R \sim 0.003 A^{1/3}$ and nuclear mass $M \sim A \times 1830$.

We mostly follow the definitions of the form factors according to the Behrens-B\"uhring formalism, as can be found in the original works \cite{Behrens1982} and Ref. \cite{Hayen2018}. For notational simplicity, however, define the following shorthand notations
\begin{align}
    {}^VF_{KLs}(\bm{q}^2) \to \left\{\begin{array}{ll}
        V_{K} & s = 0 \text{ (time)} \\
        V_{KL} & s = 1 \text{ (space)} 
    \end{array} \right.
\end{align}
and analogously for the axial vector form factors. In the Behrens-B\"uhring formalism one usually performs an expansion of the form factors in terms of $(qR)$, where the different coefficients are denoted by $F_{KLs}^{(n)}$, with $n$ the associated power of $(qR)$. In $\beta$ decay one has $(qR) \ll 1$, so that one is only concerned with $n=0$ for all form factors and $n=1$ for the dominant form factors. As such, we will leave out this additional index and denote the $n=1$ component with a prime. Finally, the form factors in the Behrens-B\"uhring formalism are typically encountered with a convolution with (parts of) the leptonic spherical wave expansion\footnote{This is discussed as the \textit{convolutional finite size} correction in Ref. \cite{Hayen2018}.}. This results in a further complication of notation, e.g., $F_{KLs}^{(n)}(\rho, k, m, n)$. In the case of allowed decays, ratios of such form factors can be calculated assuming CVC which are then evaluated directly. While such results are included in the full calculation, we do not need to introduce additional notation.

Our sign conventions, however, are slightly different from those of the Behrens-B\"uhring results. Our metric and $\gamma$ matrices follow the convention by Bjorken and Drell \cite{Bjorken1964} when specified. We take the first-class axial form factors to switch sign for $\beta^+/$EC rather than their impulse approximation expressions ($g_A \to -g_A$) as is done in Refs. \cite{Behrens1978, Behrens1982}.

\section{Coefficients and form factor decomposition}
\label{app:coefficients}

Here we report on the energy-independent factors occurring in the spectral functions as described in Sec. \ref{sec:general_expr}, and comment on the impulse approximation and consequences for, e.g., second-class current searches.

\subsection{Spectral functions}

This is a reproduction of the coefficients of the shape factor in Ref. \cite{Hayen2018} entering the formulae above, with a small caveat related to the inner radiative correction to $g_A$.

The vector coefficients, ${}^VC_i$, are as follows
\begin{subequations}
\begin{align}
    {}^VC_0 &= -\frac{233}{630}(\alpha Z)^2 - \frac{1}{5}(W_0R)^2 \mp \frac{6}{35}\alpha Z W_0 R, \\
    {}^VC_1 &= \mp \frac{13}{35}\alpha Z R + \frac{4}{15}W_0R^2, \\
    {}^VC_{-1} &= \frac{2}{15} W_0 R^2 \pm \frac{1}{70} \alpha Z R, \\
    {}^VC_2 &= -\frac{4}{15}R^2
\end{align}
\end{subequations}
where the upper (lower) sign corresponds to $\beta^-$ ($\beta^+$) decay, while the modified axial vector coefficients are
\begin{subequations}
\begin{align}
    {}^AC_0 &= -\frac{1}{5}(W_0R)^2 + \frac{4}{9}R^2\left(1 - \frac{\Lambda}{20} \right) \nonumber \\
    &+ \frac{1}{3\sqrt{3}}\frac{W_0}{A_{10}}R(\mp 2 \sqrt{2}V_{11} + 2A_1) \nonumber \\
    &\pm \frac{2}{35}\alpha Z W_0 R(1 - \Lambda) - \frac{233}{630}(\alpha Z)^2 \nonumber \\
    & + \Phi\left[\pm\frac{2}{25}\alpha Z W_0 R + \frac{51}{250}(\alpha Z)^2 \right], \label{eq:AC0}\\
    {}^AC_1 &= \frac{4\sqrt{2}}{3\sqrt{3}}R\frac{V_1}{A_{10}} + \frac{4}{9}W_0R^2\left(1-\frac{\Lambda}{10}\right) \nonumber \\
    &\mp \frac{4}{7}\alpha Z R \left(1-\frac{\Lambda}{10} \right) \pm \Phi \left[\frac{2}{25}\alpha Z R \right], \\
    {}^AC_{-1} &= -\frac{1}{3\sqrt{3}}\frac{R}{A_{10}}(\pm 2\sqrt{2}V_1 + 2A_1) - \frac{2}{45}W_0R^2(1-\Lambda) \nonumber \\
    &\mp \frac{\alpha Z R}{70}+ \Phi \left[-\frac{2}{3} W_0R^2 \pm \frac{26}{25}\alpha Z R \right], \\
    {}^AC_2 &= - \frac{4}{9}R^2\left(1-\frac{\Lambda}{10}\right),
    \label{eq:AC2}
\end{align}
\end{subequations}
where
\begin{equation}
    \Phi = \frac{\widetilde{g}_P}{g_A} \frac{1}{(2M_NR)^2} \sim \mathcal{O}(-0.1)
\end{equation}
denotes explicitly the induced pseudoscalar contribution as in Ref. \cite{Hayen2018}. Equations \ref{eq:AC0}-(\ref{eq:AC2}) use the notation we have defined in Appendix \ref{app:notation_conventions}, while the results of Ref. \cite{Hayen2018} are written using Holstein's form factors. The translation is discussed in Appendix \ref{app:comparison}. We have left out the effects of the induced Coulomb recoil corrections (i.e. $\mathcal{O}(\alpha Z/MR)$ terms) to ${}^AC_0$, which serve to renormalize the axial vector form factor as discussed in Ref. \cite{Hayen2019c, HayenGTRC}.

The subleading terms to the $\beta$ correlations discussed in the main text use a combination of the preceding ones to provide the full description. For the $\beta$-asymmetry (Eqs. (\ref{eq:f_sigma_e}) and (\ref{eq:f_2_sigma_e})), these are

\begin{subequations}
\begin{align}
    \alpha_1^0 &= \Gamma_{11} A_{10}^2 {}^AC_0 \pm 6^{-1/2}V_0A_{10}({}^AC_0 + {}^VC_0) \nonumber \\
    & \pm \frac{W_0}{2M}\left(\Gamma_{11} A_{10}^2 \pm \sqrt{\frac{2}{3}}V_0A_{10}\right) \label{eq:alpha_0} \\
    \alpha_1^1 &= \Gamma_{11}A_{10}^2\, {}^AC_1 \pm 6^{-1/2}V_0A_{10} (^AC_1+{}^VC_1) \nonumber \\
    & + \frac{\sqrt{2}R}{3}\frac{\eta_{12}}{\Lambda_1}\left[\Gamma_{11}A_{10}^2 \Delta \mp 6^{-1/2}V_0A_{10} \Delta \right. \nonumber \\
    &\left. -\sqrt{5}\Gamma_{12}A_{10}\left(V_{21} + \frac{1}{3}W_0R(V_{22} + A_{22}) \right) \right] \nonumber \\
    & \mp \frac{5}{2M}\left(\Gamma_{11}A_{10}^2 \pm \frac{7}{5}\sqrt{\frac{2}{3}}V_0A_{10}\right) \label{eq:alpha_1} \\
    \alpha_1^2 &= \Gamma_{11}A_{10}^2 {}^AC_2 \pm 6^{-1/2}V_0A_{10}({}^AC_2 + {}^VC_2).
\end{align}
\end{subequations}
Note that for each $\alpha_i$ the first line(s) contain finite size and dynamical recoil order corrections, while the last line originates from kinematical recoil corrections, discussed in Appendix \ref{app:kin_recoil}. Additionally, we define
\begin{align}
    \Delta = &- \sqrt{\frac{2}{3}}\frac{A_{11}}{A_{10}} \pm \sqrt{\frac{1}{3}}\frac{V_{11}}{A_{10}} \nonumber \\
    &+ \frac{W_0R}{9}\frac{A_{12}}{A_{10}} + \frac{2\sqrt{2}}{15}W_0R.
\end{align}

The spin-coupling coefficients are written in terms of the $\Gamma_{ij}$ factors, with $\mathcal{S}_1 = \Gamma_{11}(1)$. These follow the definitions by Weidenm\"uller and later Behrens and B\"uhring and are given by
\begin{equation}
    \Gamma_{11}(1) = \left\{\begin{array}{rl}
        \{6J(J+1)\}^{-1/2} & J \to J \\
        - \{J/6(J+1)\}^{1/2} & J \to J + 1 \\
        \{(J+1)/6J\}^{1/2} & J \to J - 1
    \end{array} \right.
\end{equation}
and
\begin{equation}
    \Gamma_{12}(1) = \left\{\begin{array}{rl}
        -\left\{\dfrac{(2J-1)(2J+3)}{30J(J+1)}\right\}^{1/2} & J \to J \\
        - \{(J+2)/10(J+1)\}^{1/2} & J \to J + 1 \\
        -\{(J-1)/10J\}^{1/2} & J \to J - 1
    \end{array} \right..
\end{equation}

Similarly, we can write down the coefficients for the $\beta-\nu$ correlation functions (Eqs. (\ref{eq:f_1_beta_nu}) and (\ref{eq:f_2_beta_nu})). The $k=1$ terms are as follows

\begin{subequations}
\begin{align}
    \tilde{\alpha}_1^0 &= V_0^2 {}^VC_0 - \frac{1}{3} A_{10}^2 {}^AC_0 - \frac{4\sqrt{2}}{9}W_0 R A_{10}^2\Delta' \nonumber \\
    & + \frac{2}{3}\frac{W_0}{M}A_{10}^2 \label{eq:alpha_0_tilde} \\
    \tilde{\alpha}_1^1 &= V_0^2 {}^VC_1 - \frac{1}{3} A_{10}^2 {}^AC_1 - \frac{2\sqrt{2}}{15}\frac{\eta_{12}}{\Lambda_1}W_0R^2 V_0^2 \nonumber \\
    &-\frac{4\sqrt{2}}{9}R A_{10}^2 \left(\frac{\eta_{12}}{\Lambda_1}\Delta'' - \Delta' \right) \nonumber \\
    & + \frac{4}{M}A_{10}^2 \\
    \tilde{\alpha}_1^2 &= V_0^2 {}^VC_2 - \frac{1}{3} A_{10}^2 {}^AC_2 + \frac{2\sqrt{2}}{15}\frac{\eta_{12}}{\Lambda_1}W_0R^2 V_0^2 \label{eq:alpha_2_tilde}
\end{align}
\end{subequations}
where again the last line is a consequence of the kinematic recoil corrections, and 
\begin{subequations}
\begin{align}
    \Delta' &= \sqrt{\frac{2}{3}}\frac{A_1}{A_{10}} \pm \sqrt{\frac{1}{3}}\frac{V_{11}}{A_{10}} - \frac{1}{5}W_0R\frac{A_{12}}{A_{10}} \nonumber \\
    &\mp \frac{1}{3}\alpha Z \left\{\frac{18\sqrt{2}}{35}+\frac{1}{3}\frac{A_{12}}{A_{10}} \right\} \\
    \Delta'' &= \sqrt{\frac{2}{3}}A_1 \mp \sqrt{\frac{1}{3}} \frac{V_{11}}{A_{10}} - \frac{1}{9}W_0R\left\{\frac{6\sqrt{2}}{5}+\frac{A_{21}}{A_{10}} \right\}.
\end{align}
\end{subequations}
Finally, the $k=2$ terms are then
\begin{subequations}
\begin{align}
    \tilde{\alpha}_2^0 &= \frac{4}{45}\nu_{12}RW_0 - \frac{2}{M_AR}\left(V_0^2-\frac{1}{3}A_{10}^2\right)\\
    \tilde{\alpha}_2^1 &= \frac{4}{45}\nu_{12}RW_e,
\end{align}
\end{subequations}
where $\nu_{12}$ is a Coulomb function of $\mathcal{O}(1+(\alpha Z)^2)$ as before \cite{Behrens1969}.

\subsection{Impulse approximation and second-class currents}

For an experimental analysis to discern new physics phenomena from Standard Model input, one needs a way of translating form factors into nuclear matrix elements, in particular for those which are not related by CVC. The usual approach follows the so-called impulse approximation, whereby the nuclear current is approximated as a coherent sum of individual, non-interacting, nucleon currents. This couples nicely with usual methods of computation consisting of some form of Slater determinants of single-particle wave functions. In practice, this translation is often done by performing a Foldy-Wouthuysen transformation \cite{Foldy1950}. Some care is required here, however, due to presence of second-class currents \cite{Wilkinson1971}. The latter has an opposite transformation under $G$-parity (i.e. $G = Ce^{i\pi I_2}$ \cite{Weinberg1958}) compared to the main currents, so that
\begin{align}
    {}^VF &\to {}^VF^I \pm {}^VF^{II} \nonumber \\
    {}^AF &\to \pm {}^AF^I + {}^AF^{II}
    \label{eq:SCC}
\end{align}
for $\beta^-$ ($\beta^+$), and where $I$ ($II$) stands for first (second) class currents. While Eq. (\ref{eq:SCC}) is general, the exact decomposition depends on the methods used and the frame in which the decomposition is performed. The following is a subset of the relevant form factors in impulse approximation in the Behrens-B\"uhring formalism, performed in the Breit frame (see Appendix \ref{app:kin_recoil}), translated from Ref. \cite{Behrens1978}
\begin{subequations}
\begin{align}
    V_0 &= g_V\mathcal{M}_F \left(1 \pm \frac{\widetilde{g}_S}{2M_N} \Delta_C \right) \label{eq:V0_full} \\
    A_{10} &= -g_A\mathcal{M}_{GT}\left(1 \mp \frac{\widetilde{g}_T}{2M_N} \Delta_C \right) \\
    A_1 &= g_A\mathcal{M}_{1} \mp \sqrt{3}\frac{\widetilde{g}_T}{2M_NR}\mathcal{M}_{GT} \\
    V_{11} &= g_V \mathcal{M}_{11} + \sqrt{6}\frac{\widetilde{g}_M}{2M_NR}\mathcal{M}_{GT}
    \label{eq:V11_full}
\end{align}
\end{subequations}
where the matrix elements are those defined in Ref. \cite{Behrens1978} with the same notation as in App. \ref{app:notation_conventions}\footnote{Note that since we defined axial vector form factors to switch signs for $\beta^+/$EC, it is the second-class contributions which change sign in Eqs. (\ref{eq:V0_full})-(\ref{eq:V11_full}).}, and
\begin{equation}
    \Delta_C = W_0 \pm \frac{6}{5}\frac{\alpha Z}{R}
\end{equation}
is the difference between the endpoint and Coulomb displacement energy. For $\beta^+$ mirror transitions, $\Delta_C$ is fairly close to zero, resulting in a decreased sensitivity.

The search for second-class currents in $\beta$ decay has a storied history \cite{Holstein1989, Holstein2018}, with initial experiments showing strong effects. The $A = 12$ isospin triplet system in particular has been an intense avenue of study through, e.g., a comparison of $\mathcal{F}t$ values. Additional complications due to nuclear structure make this comparison more complex than it appears at first sight, and subsequent experiments have found no strong evidence in favour of second-class currents. This remains the case in the study of the $\beta$-asymmetry in $^{19}$Ne \cite{Calaprice1975a, Combs2020}, which was identified as a more robust case through a measurement of the energy dependence of the asymmetry. 

When performing the ratio of $\mathcal{F}t_0$ values of the neutron and $^{19}$Ne in Sec. \ref{sec:new_physics_sensitivity}, there is additionally a contribution due to second-class currents as evidenced by Eqs. (\ref{eq:V0_full})-(\ref{eq:V11_full}). Due to the current constraints on second-class currents we do not take this into account and instead focus on scalar and tensor currents.

\section{Kinematic recoil in form factor decomposition methods}
\label{app:kin_recoil}
In the treatment of any multi-body decay with energy releases much smaller than at least one of the constituents, small recoil corrections appear, i.e. contributions of $\mathcal{O}(q/M) \ll 1$, where $q$ is the momentum transfer during the decay and $M$ in the mass of the decaying particle. In the case of $\beta$ decay, the energy released almost never exceeds 10 MeV, so that $q/M \sim 10^{-3}$ at most. At the current level of experimental precision, however, these terms are relevant. This fact is exacerbated when significant cancellations occur in the main matrix elements, so that these recoil-order effects are significantly boosted in relative precision (see Sec. \ref{sec:mirror}). Following Holstein \cite{Holstein2018}, it is useful to categorize recoil-order terms following their origin
\begin{itemize}
    \item Kinematical, $ \mathcal{O}(1) \times q/M$
    \item Dynamic, $\mathcal{O}(A) \times q/M$
    \item Coulombic, $\mathcal{O}(\alpha Z MR) \times q/M$
\end{itemize}
where $A$ is the mass number of the decaying nucleus and $R$ is the charge radius. Points two and three are contained in the proper description of the transition matrix element. Here we are mainly concerned with the first, and show how the effects are treated differently in different descriptions of nuclear $\beta$ decay at this time. 

The kinematical recoil order corrections arise in two different parts of the calculation. The first occurs in the evaluation of the nuclear current through a choice of frame. We can most easily show this in the method of Holstein, by explicitly expanding the product of lepton and nuclear currents as a set of Lorentz-scalars. In the case of a pure $J \to J$ vector transition, we can write
\begin{equation}
    i\mathcal{M} = l^\mu\langle f | V_\mu | i \rangle = a(q^2) \frac{P\cdot l}{2M}
    \label{eq:Holstein_simple_vector}
\end{equation}
where $P = p_f + p_i$ is the sum of initial and final four-momenta, and $a(q^2)$ is a general form factor. In the rest frame of the initial state, we can write
\begin{equation}
    P_\mu = (2M + E_R, -\Vec{q})
\end{equation}
where $E_R$ is the recoil energy of the final state and
\begin{equation}
    q = p_i - p_f = p + k.
\end{equation}
Taking the Hermitian square of Eq. (\ref{eq:Holstein_simple_vector}) one arrives at
\begin{equation}
    |\mathcal{M}|^2 = |a(0)|^2\left(l_0^2 + l_0\frac{\Vec{q}\cdot \Vec{l}}{M}\right)
\end{equation}
up to first order in $q/M$ and neglecting $E_R/M$. Using now the conservation of the lepton current, $\partial^\mu l_\mu = 0$, we find
\begin{equation}
    |\mathcal{M}|^2 = l_0^2|a(0)|^2\left(1 + \frac{W_0}{M}\right)
    \label{eq:Holstein_simple_recoil_lab}
\end{equation}
where $W_0$ is the energy difference between initial and final states, and $l_0^2|a(0)|^2$ represents the main transition amplitude squared. 

Moving to the Breit frame now, where $\vec{p}_i = -\vec{p}_f$ we find 
\begin{equation}
    P_\mu = (2M + X, \Vec{0})
\end{equation}
by construction, and the second term in Eq. (\ref{eq:Holstein_simple_recoil_lab}) does not appear. This is of course no problem, since one should also evaluate the lepton current in this frame. The multipole decomposition of the leptonic and hadronic currents is performed following standard methods in the Breit frame \cite{Goldberger1958, Durand1962}.
In the usual multipole decompositions \cite{Stech1964, Donnelly1975, Behrens1982}, however, one neglects the difference between lab frame and Breit frame and considers the expansion of the nuclear current correct only up to zeroth order in $\mathcal{O}(q/M)$ whether explicitly or implicitly stated. As a consequence, the results in the usual formalisms must be corrected through a Lorentz transformation from the Breit frame to the lab frame. The corrections introduced are different for different spectral functions.

A second contribution to kinematical recoil order corrections comes from the treatment of the energy integral. Rather than perform the three-body momentum integral, one sets the recoiling particle momentum to zero and instead introduces an effective correction to the transition rate
\begin{equation}
    d\Gamma \propto |\mathcal{M}|^2\left( 1 + \frac{3W_e - W_0 - 3\Vec{p}_e\cdot \hat{k}}{M} \right)
    \label{eq:phase_space_recoil_correction}
\end{equation}
where $\hat{k}$ is a unit vector in the direction of the (anti)neutrino three-momentum. Combining this result with, e.g., Eq. (\ref{eq:Holstein_simple_recoil_lab}) one then easily recovers the main kinematical recoil order corrections for vector transitions. The term proportional to $W_0/M$ cancels with Eq. (\ref{eq:Holstein_simple_recoil_lab}), and $\Vec{p}\cdot \hat{k}$ integrates to zero unless combined with similar terms in $|\mathcal{M}|^2$. To lowest order the latter is the $\beta$-$\nu$ correlation, $f_1^{\beta \nu}/f_0$ $(a_{\beta \nu})$. Performing the angular integrals one obtains finally
\begin{equation}
    {}^VR_N \approx 1 + \frac{W_e}{M}(3 - a_{\beta\nu}^{LO}).
\end{equation}
In the case of a pure vector transition one has to lowest order $a_{\beta \nu} = 1$, and one recovers the usual term, 
\begin{equation}
    {}^VR_N\approx 1 + 2\frac{W_e}{M}.
\end{equation}
Higher-order corrections and similar results for Gamow-Teller transitions can be found, e.g., in Refs. \cite{Shekhter1959, Wilkinson1982, Hayen2018, Holstein1974}.

\section{Comparison of popular formalisms}
\label{app:comparison}
When comparing to other formalisms it is important to once again take note of the fact that the form factor decomposition is non-unique (e.g. Eq. (\ref{eq:H_0_decomp})), meaning that the definition of, e.g., $\rho$ (Eq. (\ref{eq:rho_long})) is too. Thankfully, for more complex nuclei typically only two systems are in widespread use, i.e. the multipole decomposition in the Breit frame introduced by Stech and Sch\"ulke which is followed here \cite{Stech1964, Schulke1964}, Donelly and Walecka (assuming infinitely heavy nuclei) \cite{Donnelly1975, Donnelly1979}, and others, and the manifest Lorentz-invariance expansion suitable to allowed decays by Holstein \cite{Holstein1974}. While in the neutron several different works exist by a multitude of authors \cite{Ando2004, Gudkov2006, Gardner2004, Ivanov2013}, the situation there is simple enough to allow explicit spinorial calculations.

Since both approaches have been in use for several decades, compilations of comparisons have already been reported and we can be brief. The comparison between the results here using the Behrens-B\"uhring formalism and others employing the Breit frame multipole decomposition is trivial and consists only of simple prefactors. In particular, that by Donnelly and Walecka \cite{Donnelly1975, Donnelly1979, Walecka2004} which is now being used by the Jerusalem group \cite{Glick-Magid2016a} can be found in Ref. \cite{Behrens1982}. In the notation of the Jerusalem group, we can write 
\begin{subequations}
\begin{align}
    \langle J_f ||\hat{C}_J(q) || J_i \rangle &= \mathcal{C} \frac{(qR)^J}{(2L+1)!!}F_{JJ0}(q^2), \\
    \langle J_f ||\hat{L}_J(q)|| J_i \rangle &= -\mathcal{C} \left\{\frac{(qR)^{J-1}}{(2J-1)!!}\sqrt{\frac{J}{2J+1}}F_{JJ-11}(q^2) \right. \nonumber \\
    &\left. - \frac{(qR)^{J+1}}{(2J+3)!!}\sqrt{\frac{J+1}{2J+1}}F_{JJ+11}(q^2)\right\}, \\
    \langle J_f || \hat{M}_J || J_i \rangle &= \mathcal{C}\frac{(qR^J)}{(2J+1)!!}F_{JJ1}(q^2), \\
    \langle J_f || \hat{E}_J || J_i \rangle &= - \mathcal{C} \left\{ \frac{(qR)^{J-1}}{(2J-1)!!}\sqrt{\frac{J+1}{2J+1}}F_{JJ-11}(q^2) \right. \nonumber \\
    &\left. + \frac{(qR)^{J+1}}{(J2+3)!!}\sqrt{\frac{J}{2J+1}}F_{JJ+11}(q^2) \right\},
\end{align}
\end{subequations}
with $\mathcal{C} = \sqrt{\frac{2J_i+1}{4\pi}}$. Since $qR \ll 1$ in $\beta$-decay, the second and last expression can be reduced to their first term, so that there is a one-to-one translation between form factors. For the leading order $J=0$ terms this is particularly trivial. Note that because the Walecka decomposition formally occurs with infinitely heavy initial and final nuclear states -  meaning the lab frame coincides with the Breit frame where such a decomposition is justified - additional kinematic recoil corrections must included a posteriori as discussed in Appendix \ref{app:kin_recoil}. These currently do not appear to be accounted for in Ref. \cite{Glick-Magid2016a}.

The translation of Behrens-B\"uhring form factors to those of Holstein can also be found in a variety of places in the literature \cite{Holstein1974, Behrens1978, Behrens1982}. Final expressions agree perfectly if one takes into account the phase space recoil correction factor (Eq. (\ref{eq:phase_space_recoil_correction})). In the Breit frame one finds for the leading order terms \cite{Behrens1982}
\begin{subequations}
\begin{align}
    V_0 &= a + \frac{W_0}{2M}e, \\
    V_{01} &= \frac{3}{2MR}e, \\
    A_{10} &= - \left[c - \frac{W_0}{2M}d - \frac{1}{3}h \frac{\bm{q}^2}{(2M)^2} \right], \\
    V_{11} &= - \sqrt{\frac{3}{2}} \frac{b}{MR}, \\
    A_1 &= - \frac{\sqrt{3}}{2MR}\left[d + \frac{W_0}{2M}h \right], \\
    A_{12} &= - \frac{5\sqrt{2}}{4}\frac{h}{(MR)^2}.
\end{align}
\end{subequations}
When neglecting second-class currents and higher-order recoil corrections, we can make the simplifications
\begin{subequations}
\begin{align}
    V_0 &= a, \\
    A_{10} &\approx -c, \\
    A_1 &\approx \frac{\sqrt{3}}{2MR}d
\end{align}
\end{subequations}
as was done in Ref. \cite{Hayen2018} and consequently in Eqs. (\ref{eq:AC0})-(\ref{eq:AC2}). Ref. \cite{Combs2020} went a step further to make formulae easier to read by absorbing prefactors, resulting in definitions quasi-identical to those of Holstein \cite{Holstein1974}. Specifically,
\begin{align}
    F^V_0 &\equiv V_0 = a \\
    F^A_0 &\equiv A_{10} \approx -c \\
    F^V_\sigma &\equiv MR\sqrt{\frac{2}{3}}V_{11} = -b \\
    F^A_\sigma &\equiv \frac{2MR}{\sqrt{3}}A_1 \approx -d.
\end{align}
Since all but $F_0^V$ have opposite sign to Holstein's conventions, commonly quoted values such as $b/Ac, d/Ac$ remain unchanged.

\bibliography{library}

\end{document}